\documentclass[10pt,twocolumn,letterpaper]{article}

\usepackage{cvpr}              % To produce the CAMERA-READY version
\usepackage{stfloats}
\usepackage{float}
\usepackage[marginal]{footmisc} % 修改脚注样式
\usepackage[margin=1in]{geometry}
\usepackage{colortbl}
\usepackage{xcolor} 
\definecolor{best_color}{rgb}{1, 0.7, 0.7}
\definecolor{second_color}{rgb}{1, 0.85, 0.7}
\definecolor{third_color}{rgb}{1, 1, 0.7}
\newcommand{\best}{\cellcolor{best_color}}
\newcommand{\second}{\cellcolor{second_color}}
\newcommand{\third}{\cellcolor{third_color}}
\definecolor{cvprblue}{rgb}{0.21,0.49,0.74}
\usepackage[pagebackref,breaklinks,colorlinks,allcolors=cvprblue]{hyperref}

\def\paperID{14243} % *** Enter the Paper ID here
\def\confName{CVPR}
\def\confYear{2025}

\title{Kiss3DGen: Repurposing Image Diffusion Models for 3D Asset Generation}
\author{
\href{mailto:jlin695@connect.hkust-gz.edu.cn}{\textcolor{black}{Jiantao Lin}}$^{1*}$ \quad \href{mailto:xyangbk@connect.ust.hk}{\textcolor{black}{Xin Yang}}$^{1,2*}$ \quad \href{mailto:mchen221@connect.hkust-gz.edu.cn}{\textcolor{black}{Meixi Chen}}$^{1*}$ \quad Yingjie Xu$^1$ \quad Dongyu Yan$^1$ \\
Leyi Wu$^1$ \quad Xinli Xu$^1$ \quad \href{mailto:xulie@52tt.com}{\textcolor{black}{Lie Xu}}$^3$ \quad \href{mailto:zhangshunsi@52tt.com}{\textcolor{black}{Shunsi Zhang}}$^3$ \quad \href{mailto:yingcongchen@hkust-gz.edu.cn}{\textcolor{black}{Ying-Cong Chen}}$^{1,2\dagger}$ \\
$^1$HKUST(GZ) \quad $^2$HKUST \quad $^3$Guangzhou Quwan Network Technology
}

\begin{document}

\twocolumn[{
\maketitle
\begin{figure}[H]
    \vspace{-2em}
    \hsize=\textwidth
    \centering
    \includegraphics[width=1.85\linewidth]{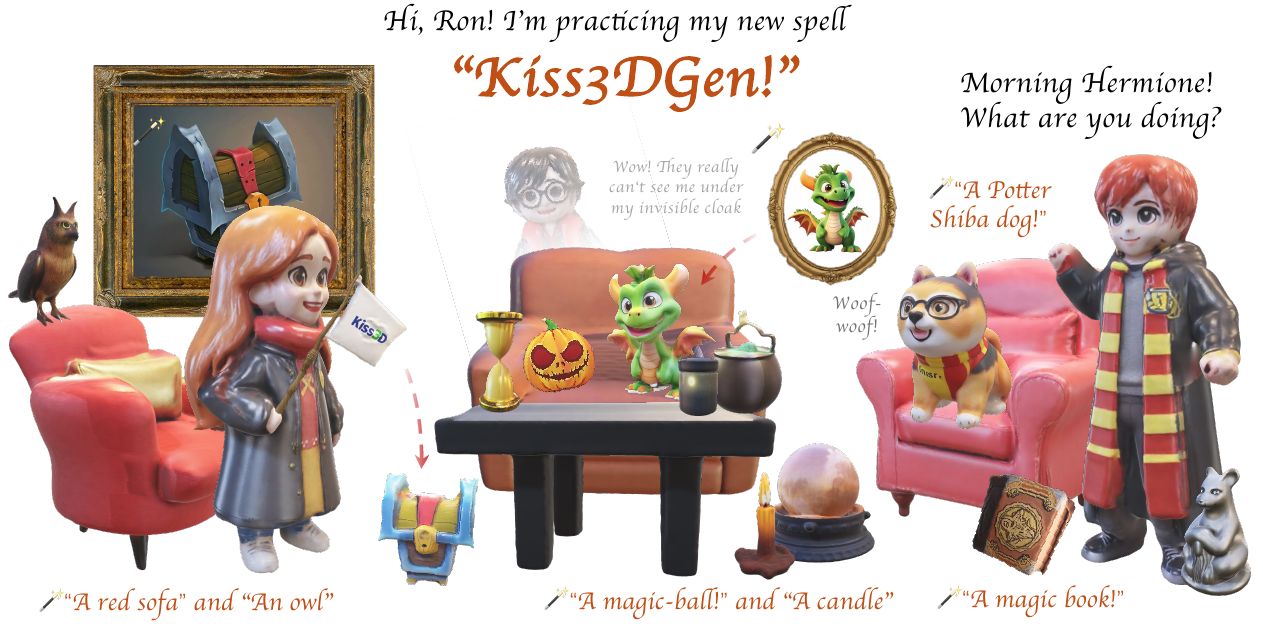}
    \vspace{-0.5em}
    \caption{\textbf{A 3D \textit{Harry Potter} scene built with Kiss3DGen.} Our proposed framework, KISS3DGen, is a unified 3D generation framework that facilitates various 3D generation tasks, including text-to-3D, image-to-3D, 3D enhancement, editing and more. Specifically, most of the assets in the figure is generated from text (captioned with abbreviated \textcolor[RGB]{178, 86, 40}{text prompts}) or image (marked by \textcolor[RGB]{176, 37, 24}{dash lines}) conditions, while the main characters (\textit{Hermoine}, \textit{Ron} and \textit{Potter}) are created using a hybrid pipeline that combines image-to-3D and text-guided mesh editing. Please zoom in for details and refer to our main paper for a more introduction.}
    % \vspace{-1em}
    \label{fig:teaser}

\end{figure}
}]

\begin{abstract}
\noindent\footnotetext{* Equal contribution \\ $\dagger$ Corresponding author}
Diffusion models have achieved great success in generating 2D images. However, the quality and generalizability of 3D content generation remain limited. State-of-the-art methods often require large-scale 3D assets for training, which are challenging to collect. In this work, we introduce \textbf{Kiss3DGen} (\textbf{K}eep \textbf{I}t \textbf{S}imple and \textbf{S}traightforward in \textbf{3D} \textbf{Gen}eration), an efficient framework for generating, editing, and enhancing 3D objects by repurposing a well-trained 2D image diffusion model for 3D generation. Specifically, we fine-tune a diffusion model to generate ''3D Bundle Image'', a tiled representation composed of multi-view images and their corresponding normal maps. The normal maps are then used to reconstruct a 3D mesh, and the multi-view images provide texture mapping, resulting in a complete 3D model. This simple method effectively transforms the 3D generation problem into a 2D image generation task, maximizing the utilization of knowledge in pretrained diffusion models. Furthermore, we demonstrate that our Kiss3DGen model is compatible with various diffusion model techniques, enabling advanced features such as 3D editing, mesh and texture enhancement, etc. Through extensive experiments, we demonstrate the effectiveness of our approach, showcasing its ability to produce high-quality 3D models efficiently. Project page: \protect\phantomsection \href{https://ltt-o.github.io/Kiss3dgen.github.io}{https://ltt-o.github.io/Kiss3dgen.github.io}.
\end{abstract} 

\section{Introduction}
\label{sec:intro}
\footnotetext{This project is supported by Quwan Technology.}
In recent years, generative models have significantly advanced the field of 3D object generation, transforming computer vision and graphics. The rapid progress of diffusion models led to remarkable improvements in synthesizing highly realistic 2D images. These developments have opened new possibilities for creating complex 3D objects, essential for applications such as virtual reality, gaming, and scientific simulations. However, extending these advancements from 2D to 3D remains challenging due to the inherent complexity of 3D geometry, the scarcity of high-quality datasets, and the high computational costs involved.

Existing 3D generation methods generally fall into two categories: optimization-based approaches and direct generation approaches. Optimization-based methods, such as DreamFusion \cite{poole2022dreamfusion}, ProlificDreamer \cite{wang2024prolificdreamer} and LucidDreamer \cite{liang2024luciddreamer}, utilize pre-trained 2D diffusion models to create 3D content. While these methods have shown promising results, they are often time-consuming during inference time because of the intensive iterative optimization required to update the 3D representation. 
% ~\cite{poole2022dreamfusion,liang2024luciddreamer,wang2024prolificdreamer}.

In contrast, direct 3D generation aims to create 3D models directly, without extensive optimization. Typical approaches include InstantMesh \cite{xu2024instantmesh}, Unique3D \cite{wu2024unique3d}, PRM \cite{ge2024prm}, Clay \cite{zhang2024clay}, Craftsman \cite{li2024craftsman}, and Direct3D \cite{wu2024direct3d}. These methods are favored in practical 3D generation systems due to their fast generation speed. However, despite their potential, these techniques rely heavily on large-scale 3D datasets, which often suffer from limitations in quality and availability. For example, Objaverse-XL \cite{deitke2024objaversexl}, the largest 3D dataset available, contains approximately 10 million samples, but around 70\% of the data is compromised by missing textures, low resolution, or poor aesthetics. This lack of high-quality data significantly hampers the effectiveness of these direct 3D generation methods. In contrast, 2D datasets such as LAION-5B \cite{schuhmann2022laion} contain billions of high-quality images, underscoring the disparity in data availability between 2D and 3D content.

Recent research has indicated that 2D diffusion models inherently possess powerful 3D priors that can be utilized for 3D understanding and generation~\cite{he2024lotus, ke2023repurposing}. However, these methods focus on producing 2.5D representations like depth or normal maps, which provide a partial view of the 3D structure and insufficient for full 3D generation. This raises an important question: Can we leverage the expressive priors learned by 2D diffusion models to create complete 3D representations?

In this work, we introduce Kiss3DGen, a simple yet effective framework for 3D asset generation. Our approach involves fine-tuning a powerful diffusion transformer model (DiT), such as Flux~\cite{blackforest2024flux}, for 3D generation tasks. Specifically, given a set of 3D objects for training, we first render each object into four distinct views along with their corresponding normal maps, forming a comprehensive representation we call the ``3D Bundle Image'', which captures both geometry and texture. Note that 3D Bundle Image is essentially an image, making it highly compatible with the existing knowledge of pretrained diffusion models, thus ensuring easy training and strong generalization capabilities.

Concretely, we use GPT-4V(ision) to generate descriptive captions for these 3D Bundle Images based on their RGB portion. These caption-image pairs are then used to fine-tune the Flux Model with Low-Rank Adaptation (LoRA)~\cite{hu2022lora}, resulting in our core model: Kiss3DGen-Base. During testing, given a user-provided caption, Kiss3DGen generates a 3D Bundle Image, which is subsequently optimized using existing mesh reconstruction approaches (e.g., NeuS~\cite{long2023wonder3d}, MeshFormer~\cite{liu2024meshformer}, ISOMER~\cite{wu2024unique3d}) produce a final 3D mesh. Thanks to the generative power of the Flux model, Kiss3DGen is capable of generating 3D content beyond the training distribution.
%(see Fig.~\cite{fig:teaser} for an example).

Since Kiss3DGen is fundamentally a diffusion model, it is naturally compatible with various diffusion-based techniques. In this work, we demonstrate this by incorporating ControlNet~\cite{zhang2023adding} to extend Kiss3DGen's capabilities, which is termed Kiss3DGen-ControlNet.  Given a 3D mesh, Kiss3DGen first renders a 3D Bundle Image, which serves as a condition for ControlNet to perform tasks such as mesh enhancement or editing tasks. Leveraging ControlNet, Kiss3DGen efficiently handles Image-to-3D generation (e.g., enhancing a coarse mesh), 3D editing (e.g., via ControlNet-Canny), and quality enhancement via ControlNet-Tile, which rebuilds low-quality 3D meshes to higher-quality.

Our contributions can be summarized as follows:

\begin{itemize}
    \item We propose Kiss3DGen, a simple yet effective approach that retargets 2D diffusion models for 3D asset generation tasks.

    % \item We extend Kiss3DGen with Kiss3DGen-ControlNet, enabling diverse functionalities such as Image-to-3D, 3D editing, and 3D quality enhancement.
    \item We show that Kiss3DGen can be seamlessly integrated with ControlNet, enabling diverse functionalities such as text-to-3D generation, image-to-3D generation, 3D editing, and 3D asset enhancement. 

    \item Extensive experiments demonstrate that the Kiss3DGen model achieves state-of-the-art performance across various tasks.
\end{itemize}

\section{Related Works}
\label{sec:related_works}
% In Sec.~\ref{sec:3d-gen}, we analyze the 3D generation frameworks based on 3D distillation approaches, multi-view models, and feed-forward 3D models. In Sec.~\ref{sec:3d-edit}, we summarize recent developments in general 3D generation tasks, such as 3D enhancement and editing.

\subsection{3D Generation}
\label{sec:3d-gen}
\noindent\textbf{Distillation-based 3D generation.} 
The rapid advancement of large-scale generative models, particularly the remarkable success of 2D diffusion models~\cite{ramesh2022hierarchical, rombach2022high, saharia2022photorealistic}, has driven significant progress in 3D reconstruction. Pioneering methods like DreamFusion~\cite{poole2022dreamfusion} and SJC~\cite{wang2023score} try to distill a 3D representation like NeRF~\cite{mildenhall2021nerf} or Gaussian Splatting~\cite{kerbl3Dgaussians} from a 2D image diffusion by a Score Distillation Sampling (SDS) loss and its variants. Follow-up distillation-based methods~\cite{wang2024prolificdreamer, lin2023magic3d, chen2023fantasia3d, melas2023realfusion, raj2023dreambooth3d, liang2024luciddreamer} attempt to improve the quality and efficiency. While these methods offer versatility and general applicability across diverse object categories, they frequently encounter challenges with convergence due to noisy and inconsistent gradient signals. This instability often results in incomplete reconstructions or artifacts, such as the ``multi-faced Janus problem''. Additionally, these methods generally demand extensive optimization time, limiting their practicality in real-world applications where speed and efficiency are essential.

% \subsection{Multi-view Generative Models}
% \label{sec:mv-diffusions}
\noindent\textbf{Multi-view generation.} 
Multi-view generation aims to generate multiple viewpoints of an object with a given image or textual prompt. Early efforts in multi-view generation models, such as MVDream~\cite{shi2023MVDream}, have successfully adapted pretrained text-to-image diffusion models~\cite{rombach2022high} to generate object-centric multi-view images. Concurrent studies~\cite{liu2023zero1to3, shi2023zero123plus, voleti2024sv3d, melaskyriazi2024im3d} explore image-conditioned multi-view generation, achieving impressive quality in multi-view outputs. However, they primarily focus on multi-view RGB generation and often overlook the challenges of 3D reconstruction. 
While some studies~\cite{wu2024unique3d, lu2024direct2} explore the separate generation of color and normal maps, another line of research achieves joint generation of both modalities with model context switcher~\cite{long2023wonder3d, li2024era3d}. The generation of normal images profoundly enhances the accuracy and quality of 3D shape formation. 
However, these approaches significantly modify the network architecture or the input-output patterns of the pretrained models and discard textural conditions, thus reducing their effectiveness for general tasks like 3D refinement and editing.
To our knowledge, no existing model unifies color and geometry generation with text conditions, a capability we believe is crucial for extending models to various 3D generation tasks, which we have achieved (see Sec.~\ref{sec:method}).

\noindent\textbf{Feedforward 3D generation.}
Beyond multi-view image generation, feedforward 3D generation refers to generating 3D representations of objects. Notably, several models within the Large Reconstruction Model (LRM) series, including MeshLRM~\cite{wei2024meshlrm}, LGM~\cite{tang2025lgm}, Instant3D~\cite{li2023instant3d}, InstantMesh~\cite{xu2024instantmesh}, and GS-LRM~\cite{GS-LRM}, use a single-image-to-multi-view generation approach to produce fixed-pose multi-view images, followed by a robust sparse-view reconstruction model to generate the final 3D assets. Distinct from the LRM series, another category of models incorporates diffusion models, such as Craftsman~\cite{li2024craftsman} and Direct3D~\cite{wu2024direct3d}, which typically follow a two-stage process: first, training a 3D variational autoencoder (VAE) to encode 3D structural information, then applying a latent diffusion model to generate 3D assets conditioned on input text or images. While these models yield high-quality results, their generalizability and robustness are limited due to constrained 3D training datasets.
In contrast, some studies leverage the strong priors of 2D diffusion models. For example, methods such as ATT3D~\cite{lorraine2023att3d} and LATTE3D~\cite{xie2024latte3d} distill 2D diffusion model priors to construct feed-forward text-to-3D generation models, though the quality of 3D assets generated in these approaches remains suboptimal. Similarly, models such as PI3D~\cite{liu2024pi3d} employ 2D diffusion priors to generate triplane representations; however, this approach modifies the original training data structure of the stable diffusion model, disrupting its intrinsic priors and significantly limiting its applicability in open-domain generation.
Kiss3DGen can seamlessly integrate with these models in multiple ways. On the one hand, these models can generate (coarse) meshes that contribute to enhancing the stability of Kiss3DGen's reconstruction phase. One the other hand, Kiss3DGen can refine and further edit these meshes with prior knowledge inherited from diffusion models. This largely improves their level of detail and overall visual quality, as well as the capacities to generate open-domain 3d assets.  

\subsection{3D Enhancement and Editing}
\label{sec:3d-edit}

3D enhancement and editing means to repair and refine initial low-quality objects, or add, delete and stylize objects using text or user interaction.
Early works like EditNeRF \cite{liu2021editing} and CLIP-NeRF \cite{wang2022clip} achieves simple part removal and colorization by feeding different codes into pretrained conditional NeRF.
For more fine-grained enhancement and editing, methods \cite{palandra2024gsedit, park2023ed, sella2023vox, haque2023instruct, kamata2023instruct} combine Instructpix2pix \cite{brooks2023instructpix2pix} and SDS to add precise text editing instructions.
However, these approaches are often time-consuming due to their optimization-based frameworks, and their implicit 3D representations are not well-suited for mesh enhancement.
DreamEditor \cite{zhuang2023dreameditor} and Focal Dreamer \cite{li2024focaldreamer} propose to use mesh-based neural field and DMTet \cite{shen2021dmtet} respectively for direct mesh optimization.
Progressive3D \cite{cheng2023progressive3d} and Focal Dreamer \cite{li2024focaldreamer} achieve convenient user interaction by spatial masking and 3D composition.
Additionally, MVEdit~\cite{mvedit2024} proposes a text-to-3D diffusion model for 3D initialization followed by a refining process guided by a 2D diffusion prior.
Coin3D \cite{dong2024coin3d} uses 3D volume adapter and  coarse proxy to aid score distillation process and achieves stronger 3D control.
Despite these advancements, such methods primarily focus on sculpting the 3D representation in a view-independent manner, often resulting in suboptimal global coherence in both geometry and texture. 
In contrast, our proposed Kiss3DGen naturally offers 3D enhancement and editing, achieving high-quality results with a streamlined pipeline.

\vspace{-0.1cm}
\section{Proposed Method}
\label{sec:method}
In this section, we present an in-depth explanation of Kiss3DGen, which repurposes a powerful diffusion transformer model (DiT) for 3D generation tasks. In Sec.~\ref{sec:Illusion-Base}, we explain how to train the base model to generate 3D Bundle Images, ultimately enabling text-to-3D generation, termed as Kiss3DGen-Base. Notably, Kiss3DGen-Base is essentially an image generation model that can be combined with many existing techniques to achieve more advanced functionalities, which will be discussed in Sec.~\ref{sec:Kiss3DGen-ControlNet}.

% We first introduce the preparation of training data and annotation processes, followed by a detailed explanation of model training in Section ~\ref{sec:Illusion-Base}. Subsequently, we discuss the combination of Illusion3D-Base with ControlNet models and the diverse applications that result from this integration in Section~\ref{sec:Illusion-ControlNet}. An overview of the framework is illustrated in Fig~\ref{fig:framework}.
\begin{figure}[t]
    \centering
    \includegraphics[trim= 0 20 0 0, width=1.0\linewidth]{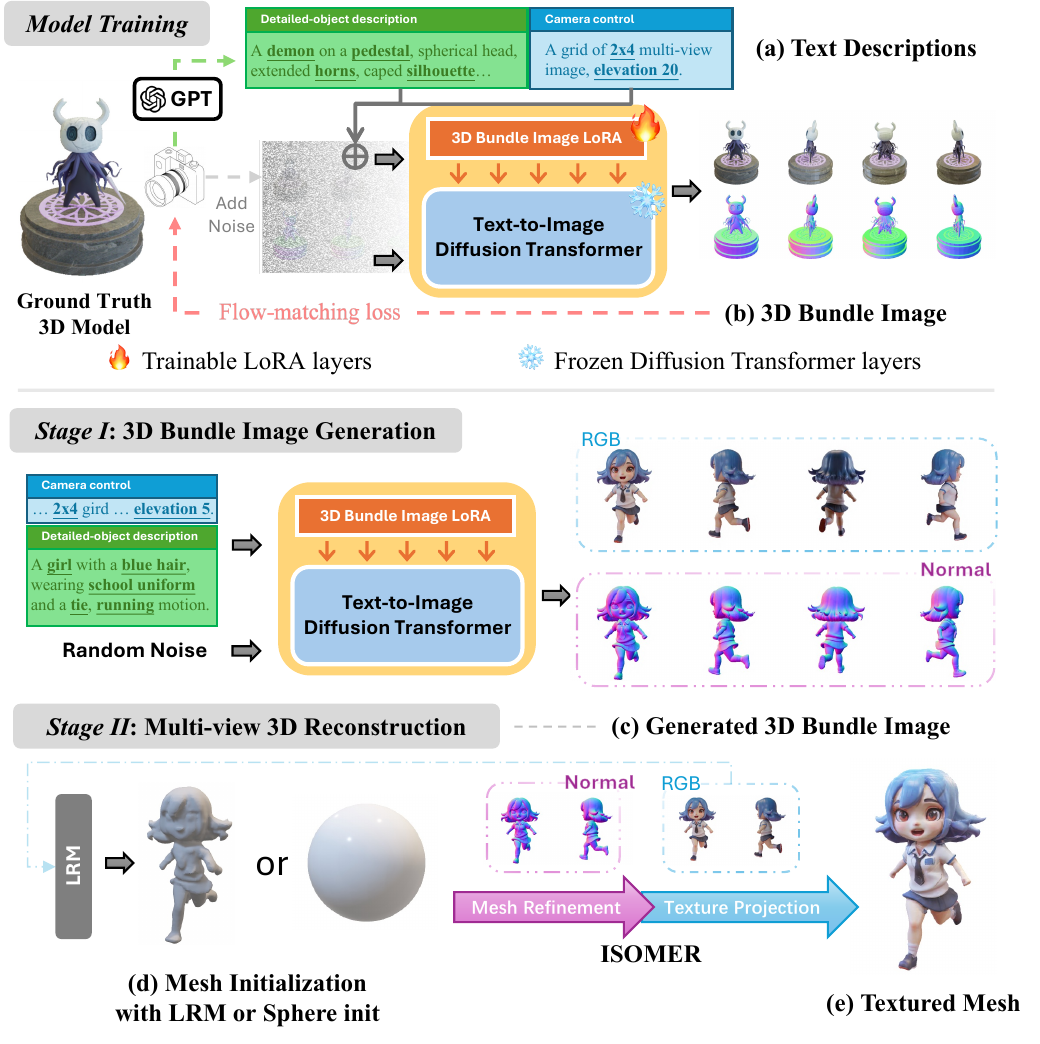}
    \caption{\textbf{The overview of our text-to-3D training and generation framework.} 
    In this work, we curate a high-quality text-3D dataset, then train a LoRA~\cite{hu2022lora} layer for text to 3D bundle image (Sec.~\ref{sec:method}) generation upon a pretrained text-to-image diffusion transformer model with flow matching. Our framework generates 3D assets with text condition in two stages: the 3D-Bundle-Image generation (\textit{Stage I}) and the 3D reconstruction (\textit{Stage II}). In \textit{Stage I}, we generate 3D bundle image with our Kiss3DGen base model guided by text prompts. In \textit{Stage II}, we reconstruct the geometry and texture of the 3D asset via LRM~\cite{xu2024instantmesh, hong2024lrmlargereconstructionmodel} or sphere initialization followed by optimization-based mesh refinement and texture projection approach, i.e., ISOMER~\cite{wu2024unique3d}. Zoom in for details.}
    \label{fig:framework}
    \vspace{-1em}
\end{figure}

\subsection{Kiss3DGen-Base}
\label{sec:Illusion-Base}
Kiss3DGen-Base is designed to generate high-quality 3D Bundle Images that encapsulate both the geometry and texture information of 3D objects. Then, we leverage ISOMER \cite{wu2024unique3d} to generate textured 3D meshes with these 3D Bundle Images. The overarching design principle is to align the 3D Bundle Image with the prior distribution of pre-trained image diffusion models, thereby preserving the original generative capabilities of the pre-trained model to the greatest extent possible.

% \paragraph{3D Bundle Image}
\noindent \textbf{3D Bundle Image.}~
Given a 3D object, we render it into four distinct views along with their corresponding normal maps, creating a comprehensive multi-view representation called the 3D Bundle Image, as shown in Fig. \ref{fig:framework}(b). This approach is based on two key insights. First, the combination of normal and RGB images from multiple views capture all necessary information to form a complete 3D object, which is then converted into a 3D mesh with ISOMER \cite{wu2024unique3d}. Second, since the 3D Bundle Image is essentially a 2D representation, it naturally aligns with the prior information in pre-trained 2D diffusion models, allowing us to leverage their generative capabilities without altering the input-output structure. This ensures the model effectively integrates and utilizes the learned priors. 

% \paragraph{Model}
\noindent \textbf{Model.}~
Learning the 3D Bundle Image is a non-trivial task, as the model must capture the relationships between different views and the correspondences between RGB images and normal maps, which are inherently complicated. Given the spatial distances between RGB images and normal maps, modeling long-range dependencies is crucial. To address this, we employ a DiT model, i.e., Flux~\cite{blackforest2024flux}, whose attention blocks are particularly effective at capturing these long-range dependencies, ensuring coherent multi-view and cross-modal relationships are properly modeled.

% \paragraph{Captioning}
\noindent \textbf{Captioning.}~
To enhance the training process and leverage text-image correspondence, we generate descriptive captions for each 3D Bundle Image. We use GPT-4V to provide detailed captions that describe the content of each Bundle Image, including visual attributes such as color, shape, and surface properties. These captions help encode semantic information about the 3D objects, providing an additional supervisory signal during training that ensures the model learns to associate textual descriptions with specific geometric and visual features. An example caption can be found in Fig. \ref{fig:framework}(a). 

\noindent \textbf{Training and Inference.}~
With the 3D bundle images and the captions, we train a LoRA to retarget the pretrained Flux~\cite{blackforest2024flux} model to generate the 3D Bundle Image (Fig.~\ref{fig:framework}(b)). As a result, the model can generate a 3D Bundle Image from text prompt, producing a set of image and normal map pairs from four distinct viewpoints. 
% Subsequently, we employ ISOMER~\cite{wu2024unique3d} to optimize the mesh, which can be initialized either as a sphere or using LRM~\cite{xu2024instantmesh}, as illustrated in Fig.~\ref{fig:framework}. This process facilitates Text-to-3D generation, enabling the creation of 3D models directly from text prompts.
Then, we can employ ISOMER \cite{wu2024unique3d} to optimize a mesh, which could be initialized  with a sphere, as shown in Fig.~\ref{fig:framework}. In practice, we also found that initializing the mesh with LRM \cite{xu2024instantmesh} is more robust with a bit more inference time. Thus we adopt this strategy in this paper. A detailed study will be shown in the supplementary file. 
This process enables Text-to-3D Generation.

\vspace{-0.1cm}
\subsection{Kiss3DGen-ControlNet}
\label{sec:Kiss3DGen-ControlNet}
To extend the capabilities of Kiss3DGen beyond direct generation, we introduce Kiss3DGen-ControlNet, which incorporates ControlNet~\cite{zhang2023adding} to handle a variety of 3D-related tasks such as enhancement, editing, and image-to-3D generation. In this section, we start with 3D enhancement as a basic application and expand into multiple use cases, providing examples of how Kiss3DGen-ControlNet can be applied for the aforementioned tasks. It should be noted that its potential usages are far more extensive.

% \begin{figure}[t]
%     \centering
%     \includegraphics[trim= 50 20 10 0, width=1.0\linewidth]{sec/figures/image_to_3d.pdf}
%     \caption{\textbf{A general image-to-3D pipeline with Kiss3DGen 3D enhancement.} In order to achieve high-quality image-to-3D generation, we incorporate the existing image-to-3D pipeline~\cite{xu2024instantmesh} with our general 3D enhancement pipeline. Please Zoom in for details.}
%     \label{fig:image-to-3d}
% \end{figure}

\begin{figure}[t]
    \centering
    \includegraphics[trim=20 20 15 0, width=1.0\linewidth]{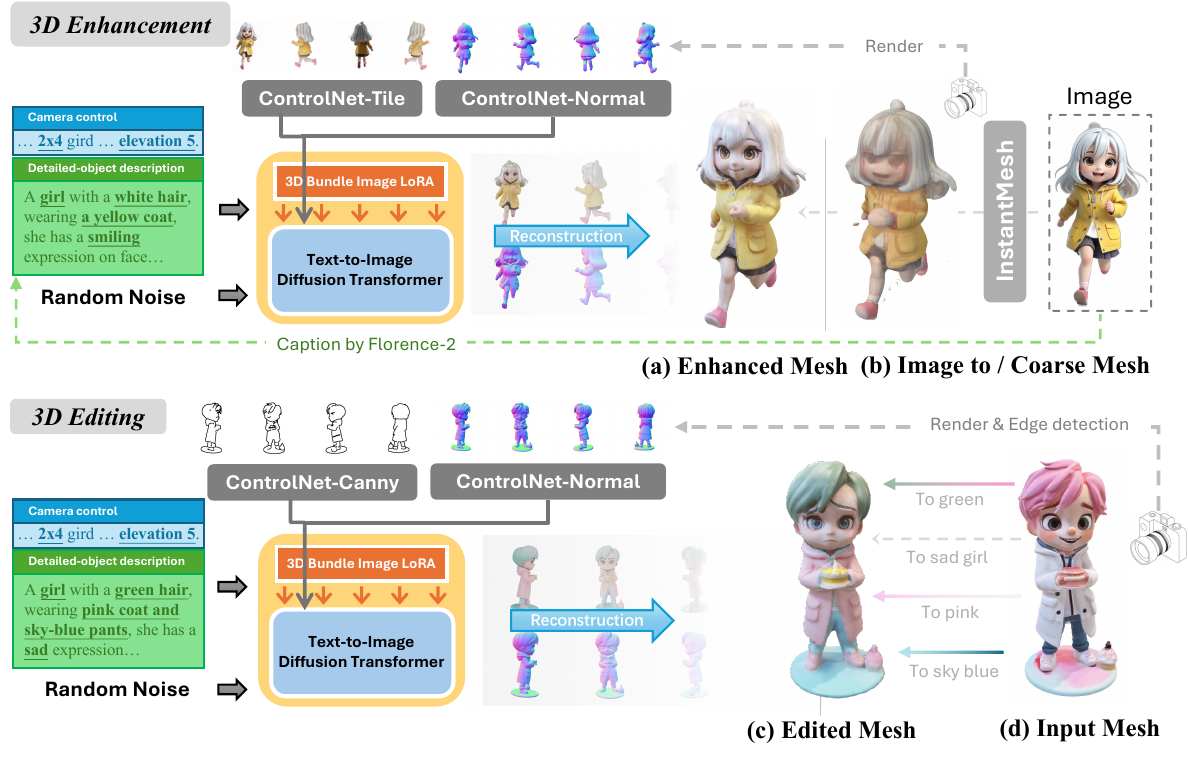}
    \caption{\textbf{3D enhancement and editing with Kiss3DGen.} In order to achieve high-quality image-to-3D generation, we incorporate the existing image-to-3D pipeline~\cite{xu2024instantmesh} with our general 3D enhancement pipeline. Please zoom in for details.}
    \label{fig:3d_refinement_and_editing}
    \vspace{-1.5em}
\end{figure}

\noindent \textbf{3D Enhancement.}
Given a low-quality mesh (e.g., Fig.~\ref{fig:3d_refinement_and_editing}(b)), which may suffer from geometry lacking detail or overly blurred textures, we can render it into a 3D Bundle Image and then process it through the ControlNet-Tile and ControlNet-Normal model within Kiss3DGen. Originally, ControlNet-Tile was used for super-resolution of images, preserving color and content while leveraging diffusion model priors to enhance details. We found that this approach is also suitable to enhance our 3D Bundle Images, including the RGB images and normal maps. Note that 3D enhancement requires retaining the semantic integrity of the original mesh as much as possible. To facilitate this, we use the Florence-2~\cite{xiao2023florence} to generate captions for the RGB portion of the 3D Bundle Image.
With the enhanced 3D Bundle Images, we use ISOMER \cite{wu2024unique3d} to further refine the input mesh. The enhanced RGB images and normal maps enrich the details of texture and geometry respectively.

It should be noted that incorporating ControlNet can limit changes to the edited model, potentially constraining improvements in geometry and texture details. To address this, we introduce two hyperparameters, $\lambda_1$ and $\lambda_2$. $\lambda_1 \in [0,1]$ represents the ControlNet Strength, determining how strongly the ControlNet Branch adds to the original network, i.e., $y=F(x) + \lambda_1 F_c(c) $, where $F(x)$ is the feature of main branch, $F_c(c)$ is the feature of the control branch. $\lambda_2 \in [0,1]$ represents the fraction of diffusion steps during which ControlNet is active. Specifically, ControlNet is applied from step 0 to $\lambda_2 T$, and is not used from step $\lambda_2 T + 1$ to $T$. Empirically, $\lambda_1$ values between 0.05 and 0.8, and $\lambda_2$ values between 0.1 and 0.7, yield good results. In most experiments, we set $\lambda_1=0.6$ and $\lambda_2=0.3$ for a balance between enhancement and flexibility.

\noindent \textbf{3D Editing.}~ 
Similar to the 3D enhancement mentioned above, by decreasing the weight of ControlNet and allowing users to provide customized captions, Kiss3DGen-ControlNet can effectively perform 3D editing. As shown in Fig.~\ref{fig:3d_refinement_and_editing}(c, d), this pipeline allows users to alter specific attributes of the 3D object, such as shape or texture, while maintaining overall coherence with the original model. Note that this is not the only way to achieve 3D editing. We have tried to convert the 3D Bundle Image with Canny operation, then apply ControlNet-Canny, or simply use SDEdit to edit the 3D Bundle Image; both work well. In this paper, we simply adopt ControlNet-Canny and ControlNet-Normal, then apply $\lambda_1=0.3$, $\lambda_2=0.5$. 

\noindent \textbf{Image-to-3D Generation.}~ 
Kiss3DGen-ControlNet supports Image-to-3D Generation. By using existing methods (e.g., InstantMesh~\cite{xu2024instantmesh}) to generate a coarse mesh from a given 2D image, Kiss3DGen-ControlNet can refine this initial output, transforming it from a low-quality, rough mesh into a high-quality 3D model. 
In Fig.~\ref{fig:3d_refinement_and_editing}(a, b), we demonstrate an example of image-to-3D generation by reusing our 3D enhancement pipeline, where we further replace one of the RGB views in the rendered 3D Bundle Image with the input image.
This two-stage approach allows for the efficient transformation of 2D inputs into detailed 3D objects, utilizing the enhancement capabilities to improve mesh quality.

% % \paragraph{Advanced Text-to-3D Generation}
% \noindent \textbf{Advanced Text-to-3D Generation}~
% While Kiss3DGen-Base can already achieve 

% \begin{figure*}[t]
%     \centering
%     \includegraphics[trim= 10 0 30 0, width=1.0\linewidth]{sec/figures/3d_editing.pdf}
%     \caption{\textbf{A general 3D editing pipeline with Kiss3DGen.} By simply apply pretrained ControlNet~\cite{zhang2023adding} to our Kiss3DGen model, it enables high quality 3D editing without further fine-tuning. Specifically, we adopt the ControlNet-Canny for shape control and ControlNet-Normal for geometry control. Please zoom in for details.}
%     \label{fig:3d-editing}
% \end{figure*}

% Because our Kiss3DGen model shares not only most weights but also the same inference pipeline with the pretrained DiT model, it is naturally compatible with most existing plugins and adapters, e.g., ControlNets~\cite{zhang2023adding}, that are built for the pretrained model without training. Therefore, the Kiss3DGen can be adapted to various generation tasks, such as 3D editing or image-to-3D generation. 

% \noindent \textbf{Kiss3DGen for 3D Editing}~
% For 3D editing.

\section{Experiments}
\label{sec:experiments}
\subsection{Dataset}
\label{sec:dataset}
Due to the inconsistent quality of the original Objaverse dataset \citep{deitke2023objaverse}, we initially excluded objects that lacked texture maps or had low polygon counts. We then conducted meticulous manual curation to remove low-quality samples, such as incomplete objects, scanned flat surfaces, and large-scale scene samples. Additionally, given the irregular orientations of objects in the Objaverse dataset, we manually annotated and corrected the front-facing angle for each object. This rigorous refinement process resulted in a collection of 147k high-quality objects, which we used to train the Kiss3DGen-Base model.
In addition, we curated a specialized dataset of 4k high-quality 3D cartoon-style human body models from the internet. This dataset was used to train the Kiss3DGen-Doll model, specifically designed to support cartoon-style character generation. 
To render these objects, we used Blender with a camera distance set to 4.5 units and a field of view (FoV) of 30 degrees. For each object, we used Blender to render four views separated by 90 degrees in azimuth, with the first view aligned to the front-facing angle of the object. The elevation is fixed as 5 degree.  Both RGB and Normal maps were rendered for each view at a resolution of $512\times512$. The four view images were then used as inputs for 3D-aware caption annotation in GPT-4V. 

\vspace{-0.1cm}
\subsection{Implementation Details}
For the training of Kiss3DGen-Base, we utilized FLUX.1-dev~\cite{blackforest2024flux} as our base model, training it with Low-Rank Adaptation (LoRA)~\cite{hu2022lora} on 8 NVIDIA A800 80GB GPUs for 3 days, completing 16 epochs with a batch size of 4. The Adam optimizer was employed with a fixed learning rate of $8 \times 10^{-4}$, and training was conducted with bf16 precision. The LoRA rank (network dimension) was set to 128.

%  In (b), we extend our framework for mesh re-texturing, which shows comparable performance against the baseline.
\vspace{-0.1cm}
\subsection{Evaluation}
\label{sec:evaluation}
% \vspace{-0.1cm}
For evaluation dataset, we conducted quantitative comparisons using the Google Scanned Objects (GSO) dataset~\citep{downs2022google}.
For the \textit{text-to-3D} evaluation, we randomly sampled 100 objects, rendering four orthogonal views per object and then annotated using GPT-4V to generate text inputs for the test set. For the \textit{image-to-3D} evaluation, we sampled 200 objects and rendered a single front-facing view for each, serving as the input for quantitative assessment.

For the evaluation protocol, we assessed alignment with text descriptions and the visual quality of the generated results. For the \textit{text-to-3D} task, without precise 3D ground truth, we measured alignment using CLIP score~\cite{radford2021learning} by rendering four orthogonal views of each generated object. Also, we used Q-Align~\cite{wu2023qalign}, a large multi-modal model, to evaluate the quality and aesthetics of the rendered images.
For the image-to-3D task, we evaluated both 2D visual quality and 3D geometric quality. For the 2D visual evaluation, we rendered novel views from the generation 3D mesh and compared them to the ground truth views using PSNR, SSIM, and LPIPS as metrics.
For the 3D geometric evaluation, we focused on comparing the generation 3D meshes to the ground truth. First, we aligned the coordinate systems of the generated and ground truth meshes. Next, we repositioned and rescaled all meshes into a cube of size $[-1, 1]^3$. We reported Chamfer Distance (CD) and F-Score (FS) at a threshold of 0.1, calculated by uniformly sampling 16K points from the mesh surfaces.

\begin{table}[]
\caption{The quantitative comparison with MVDream in \textbf{text-to-multi-view synthesis} shows that our method outperforms MVDream~\cite{shi2023MVDream} by a large margin. Notably, our method even surpasses the ``Real Data'' in both “Quality” and “Aesthetic” metrics, where “Real Data” refers to the multiple views used to generate text by GPT-4V.}
\label{tab:text-to-multiview}
\centering
\resizebox{\linewidth}{!}{
\begin{tabular}{ccccc}
\hline
    Method         & Dataset size      & CLIP$\uparrow$      & Quality$\uparrow$     & Aesthetic$\uparrow$  \\ \hline
    Real Data   & N/A                & \best{0.884}        & \second{3.138}        & \second{1.911}            \\
    MVDream        & 350K               & \third{0.809}       & 2.509                 & 1.526                   \\ 
    Ours-Base      & 147K               & \second{0.844}      & \best{3.248}          & \best{1.94}             \\ 
    Ours-50K       & 50K                & 0.804               & \third{2.972}         & \third{1.879}          \\ \hline
\end{tabular}
}
\vspace{-1em}
\end{table}

\begin{table}[]
\caption{Quantitative comparison of \textbf{text-to-3D generation} results after rendering, evaluated in terms of CLIP-score, Quality, and Aesthetic metrics. Our method outperforms 3DTopia~\cite{hong20243dtopia}, Direct2.5~\cite{lu2024direct2}, and Hunyuan3D-1.0~\cite{yang2024hunyuan3d} across all metrics, indicating a significant improvement in alignment with textual descriptions and visual quality.}
\label{tab:text-to-3D}
\centering
\resizebox{\linewidth}{!}{
\begin{tabular}{ccccc}
\hline
    Method        & Dataset size & CLIP$\uparrow$      & Quality$\uparrow$     & Aesthetic$\uparrow$  \\ \hline
    3DTopia       & 320k          & 0.694               & 2.145                 & \third{1.538}          \\          
    Direct2.5     & 500k          & 0.773               & 2.158                 & 1.459          \\
    Hunyuan3D-1.0 & N/A           & \third{0.792}       & \third{2.517}         & 1.504          \\
    Ours-Base     & 147k          & \best{0.837}        & \second{2.700}        & \best{1.800}           \\
    Ours-50K      & 50k           & \second{0.804}      & \best{2.716}          & \second{1.601}                \\ \hline
\end{tabular}
}
\end{table}

% \vspace{-0.2cm}
\begin{table}[]
\caption{Quantitative comparison of \textbf{image-to-3D generation} methods across multiple metrics. Since CraftsMan generates only geometry without textures, the comparison with this method is limited to 3D geometry metrics (Chamfer Distance (CD) and F-Score (FS)).}
\label{tab:image-to-3D}
\centering
\resizebox{\linewidth}{!}{
\begin{tabular}{ccccccc}
\hline
    Method        & Dataset size  & CD$\downarrow$      & FS$\uparrow$     & PSNR$\uparrow$    & SSIM$\uparrow$  & LPIPS$\downarrow$     \\ \hline
    CraftsMan     & 170k           & 0.178               & 0.739            & N/A               & N/A             & N/A                   \\          
    Unique3D      & 50k            & 0.217               & 0.654            & \third{19.237}    & \second{0.898}  & \third{0.127}                 \\
    Hunyuan3D-1.0 & N/A            & \third{0.153}       & \second{0.768}   & 16.652            & \third{0.885}   & \second{0.123}                 \\
    Ours-Base     & 147k           & \best{0.149}        & \best{0.769}     & \best{20.348}     & \best{0.902}    & \best{0.116}          \\
    Ours-50K      & 50k            & \second{0.151}      & \third{0.766}    & \second{20.215}   & 0.884           & 0.131                 \\ \hline
\end{tabular}
}
\vspace{-1em}
\end{table}
% \vspace{0.3cm}
\subsection{Comparison with State-of-the-Art Methods}
In this section, we compare our approach with several state-of-the-art methods across three tasks: text-to-multi-view synthesis, text-to-3D generation, and image-to-3D generation.
For text-to-multi-view synthesis, we compare with MVDream~\cite{shi2023MVDream}, which introduces a multi-view attention mechanism within its model to facilitate multi-view information interaction. This approach aims to improve consistency across generated views by allowing information to be shared among them.
For text-to-3D generation, we compare our method with three recent approaches—3DTopia~\cite{hong20243dtopia}, Direct2.5~\cite{lu2024direct2}, and Hunyuan3D-1.0~\cite{yang2024hunyuan3d}—that claim capabilities in text-to-3D generation. 3DTopia trains a latent diffusion model to generate Tri-plane representations for 3D object synthesis. Direct2.5 takes a different approach by utilizing two separate diffusion models to generate Normal maps and corresponding RGB maps, which are then used to reconstruct the geometry and apply textures. In contrast, Hunyuan3D-1.0 follows large reconstruction model (LRM) that first generates a single image from text and then expands this into six multi-view images for sparse-view reconstruction.
For image-to-3D generation, we compare the Hunyuan3D-1.0, the diffusion-based model CraftsMan~\cite{li2024craftsman}, and Unique3D~\cite{wu2024unique3d}, a two stage generation method similar to ours.
% \textcolor{blue}{current state-of-the-art in optimization-based approach method.}
% an optimization-based approach representing the current state-of-the-art.

\begin{figure}
    \centering    \includegraphics[width=1.0\linewidth]{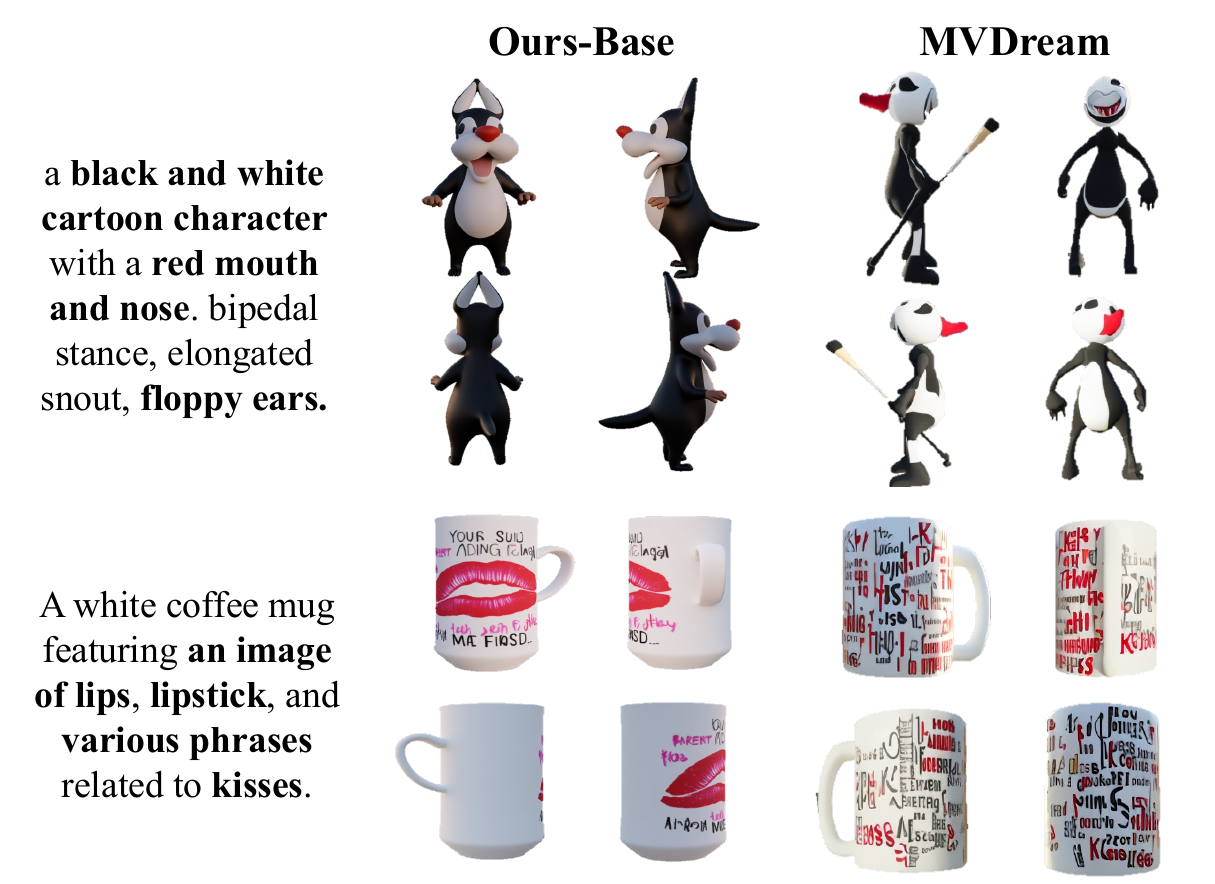}
    \caption{\textbf{Qualitative comparisons with MVDream~\cite{shi2023MVDream} in \textcolor{blue}{text-to-multiview} generation.} In comparison, our method produces significantly better results in both text-image alignment and geometric coherence.}
    \label{fig:text_to_multiview}
    \vspace{-2em}
\end{figure}

\noindent\textbf{Text-to-Multi-View Synthesis.} As shown in Tab.~\ref{tab:text-to-multiview}, we conducted a quantitative evaluation comparing our method to MVDream. Our method demonstrates a substantial improvement over MVDream across all metrics, indicating enhanced consistency and quality across multiple views. Notably, our approach outperforms the “Real Data” in both “Quality” and “Aesthetic” scores, which may seem counter-intuitive. However, this makes sense considering that our model inherits a lot of knowledge from Flux model which tends to produce high-quality images. Furthermore, even the model trained on a reduced dataset of just 50K samples achieves competitive results, indicating that our approach is highly data-efficient and can yield strong performance with limited training data. Also, we performed qualitative comparisons in Fig.~\ref{fig:text_to_multiview} which illustrate the superior multi-view coherence and realism achieved by our approach.

\begin{figure}
    \centering
    \includegraphics[width=\linewidth]{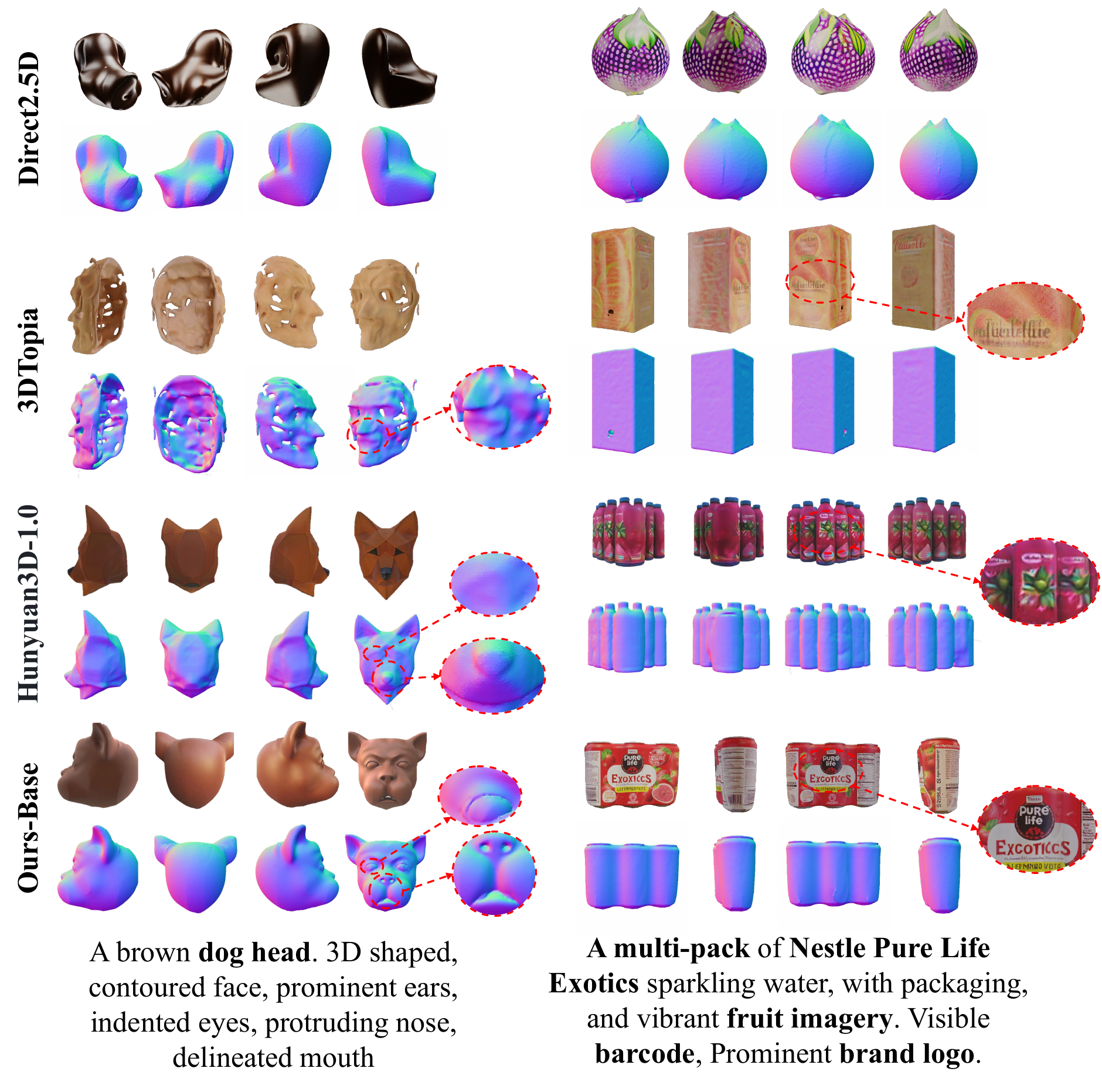}
    \vspace{-2em}
    \caption{\textbf{Qualitative comparisons with state-of-the-art methods for \textcolor{blue}{text-to-3D} generation.} It demonstrates that Kiss3DGen achieves the highest quality 3D mesh, delivering more accurate texture generation from the input prompts compared to others.}
    \label{fig:text_to_3D}
    \vspace{-0.5em}
\end{figure}

\begin{figure}
    \centering
    \includegraphics[trim= 30 20 20 20, width=0.9\linewidth]{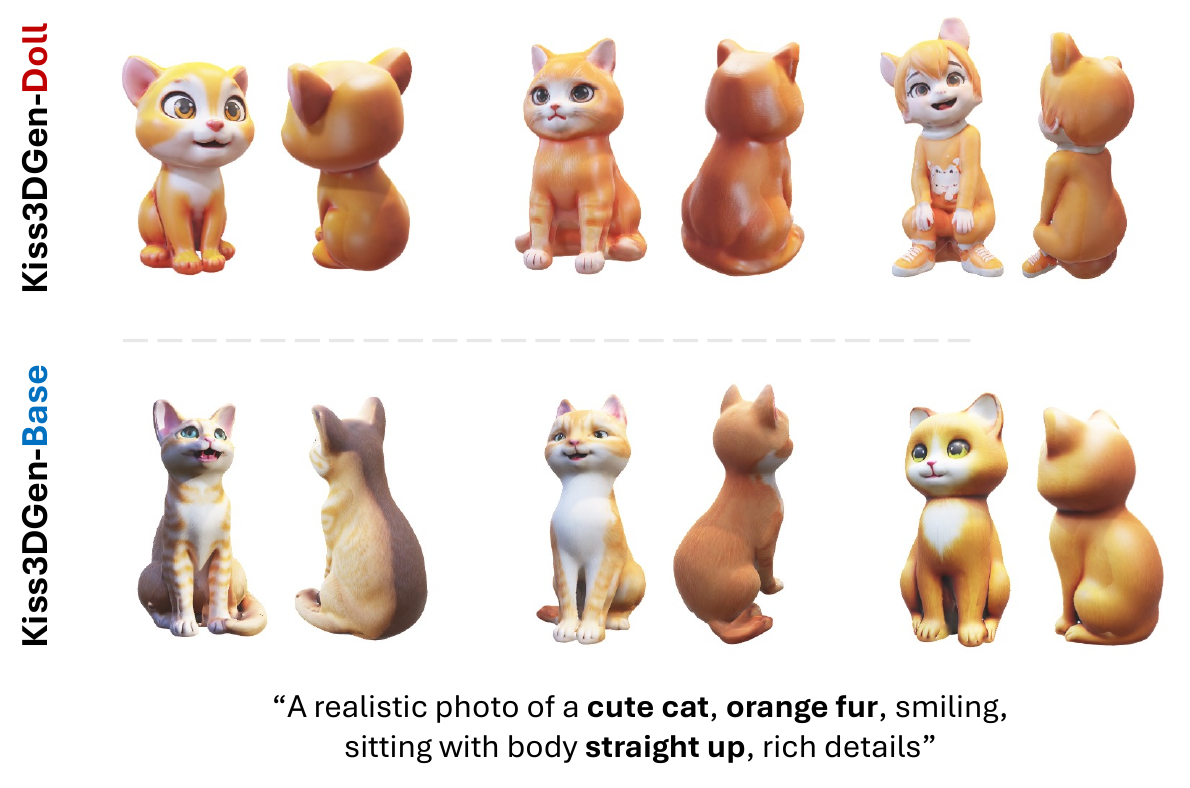}
    \caption{Text-to-3D comparison between our Base and Doll models (Sec.~\ref{sec:dataset}). Each model generates different results with different seeds. All images are rendered from 3D mesh.}
    \label{fig:base-vs-doll}
    \vspace{-1em}
\end{figure}

\begin{figure}[t]
    \centering
    % \vspace{-1em}
    \includegraphics[trim= 5 30 20 0, width=1.0\linewidth]{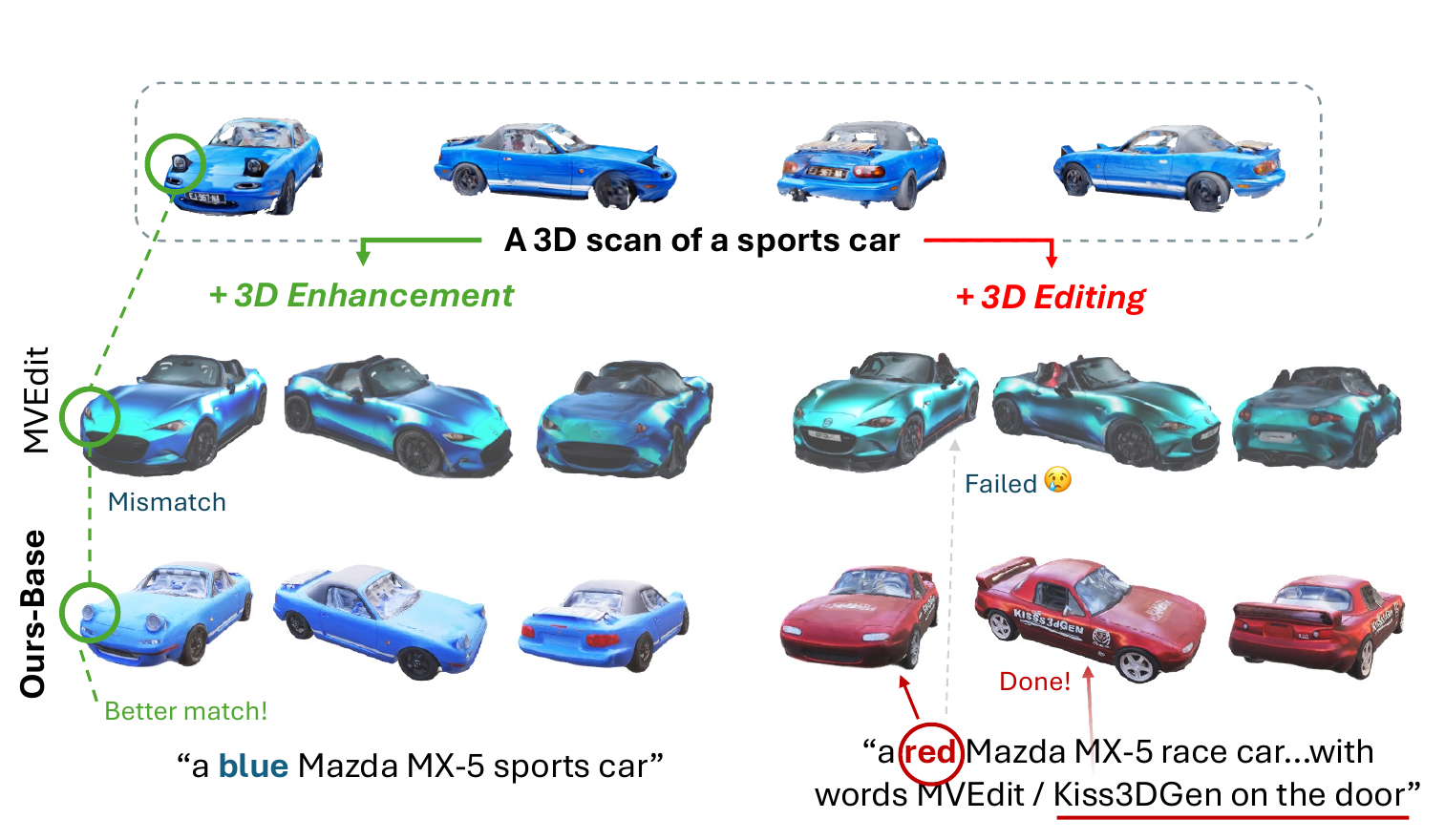}
     \caption{\textbf{Qualitative comparison on 3D mesh Enhancement and Editing with MVEdit~\cite{mvedit2024}.} Our results (line 2) maintain much better consistency with the input mesh in enhancement and better align with the given text in editing.}
    \label{fig:enhance_compare}
    \vspace{-1em}
\end{figure}

\noindent\textbf{Text-to-3D Generation.} As shown in Tab.~\ref{tab:text-to-3D}, we quantitatively compared our approach with 3DTopia, Direct2.5, and Hunyuan3D-1.0 in terms of alignment with text, visual quality, and aesthetic appeal. Our method consistently outperforms the others across all metrics, particularly in CLIP-score, which reflects improved alignment with the input text. Furthermore, we achieve higher Quality and Aesthetic scores, showcasing our method’s ability to generate not only accurate but also visually pleasing 3D representations. \textcolor{black}{Noticed that these metrics are calculated by rendering the generated 3D results into four orthogonal views.} In addition, qualitative comparisons are provided in Fig.~\ref{fig:text_to_3D}. In Fig.~\ref{fig:base-vs-doll}, we show both our Base and Doll models can serve high-quality text-to-3D generation, even for cases that are out of the training domain. We also conducted qualitative comparisons on 3D mesh enhancement and editing with MVEdit~\cite{mvedit2024} in Fig.~\ref{fig:enhance_compare}, demonstrating our ability to perform more precise mesh enhancement or editing.
% \vspace{-0.2cm}

\begin{figure}
    \centering
    \includegraphics[width=1.0\linewidth]{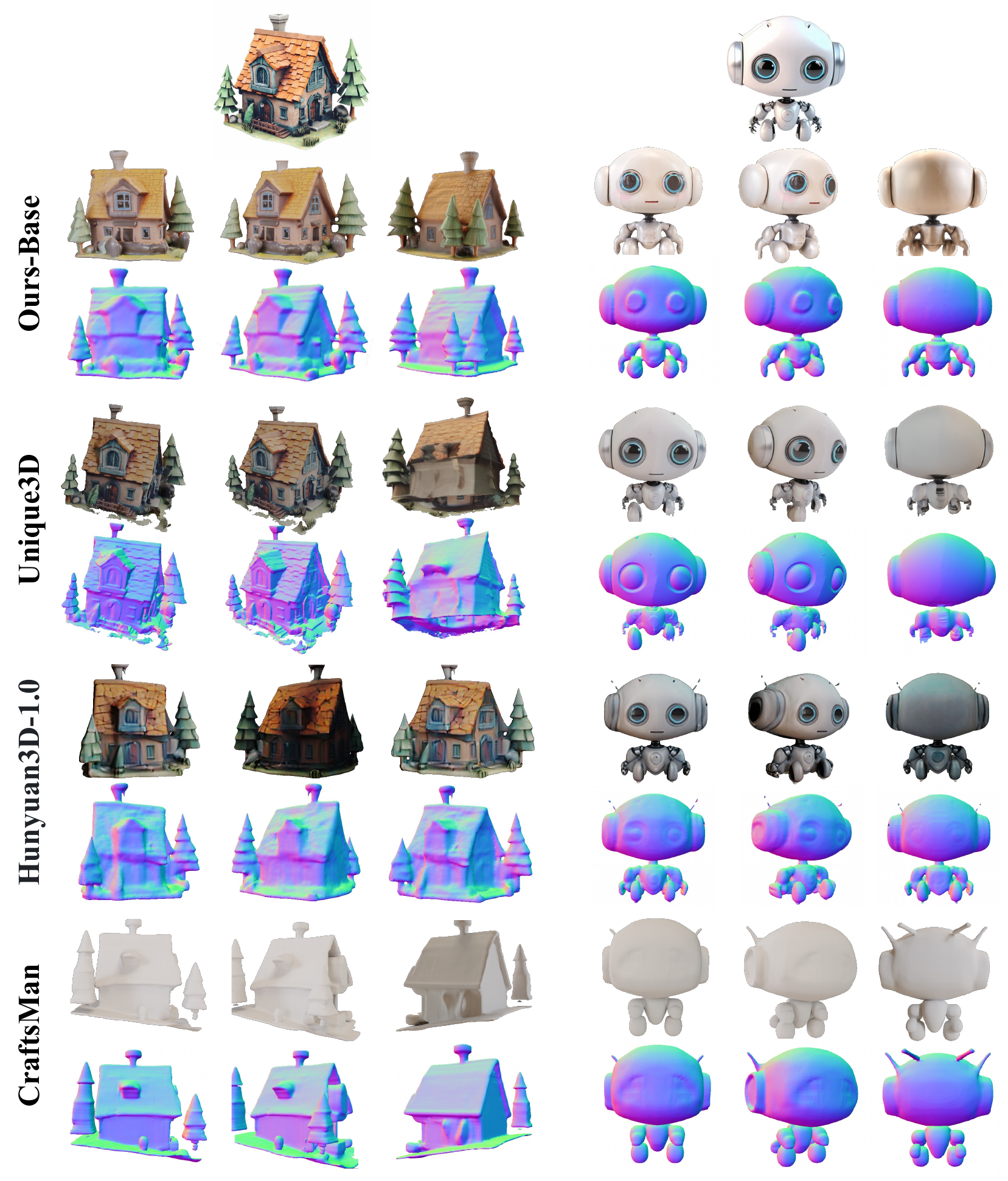}
    \caption{\textbf{Qualitative comparisons with state-of-the-art methods for \textcolor{blue}{image-to-3D} generation.} We achieves the highest quality 3D mesh, delivering more accurate and realistic texture generation from input images compared to other models.}
    \vspace{-2em}
    \label{fig:image_to_3D}
\end{figure}

\noindent\textbf{Image to 3D generation}
In Tab.~\ref{tab:image-to-3D}, we present a quantitative comparison of various image-to-3D generation methods across multiple metrics. Our approach demonstrates superior performance, outperforming other methods in Chamfer Distance (CD), F-Score (FS), PSNR, SSIM, and LPIPS scores. Notably, our model achieves strong results even with a reduced training dataset (Ours-50K), indicating its efficiency and robustness in generating high-quality 3D representations from 2D images.
In addition to these quantitative results, Fig.~\ref{fig:image_to_3D} provides qualitative comparisons that highlight the improved texture fidelity, structural coherence, and overall visual quality achieved by our approach. 

\subsection{Ablation Study}
\label{sec:ablation}
\begin{figure}[h]
    \centering
    \includegraphics[width=0.5\textwidth]{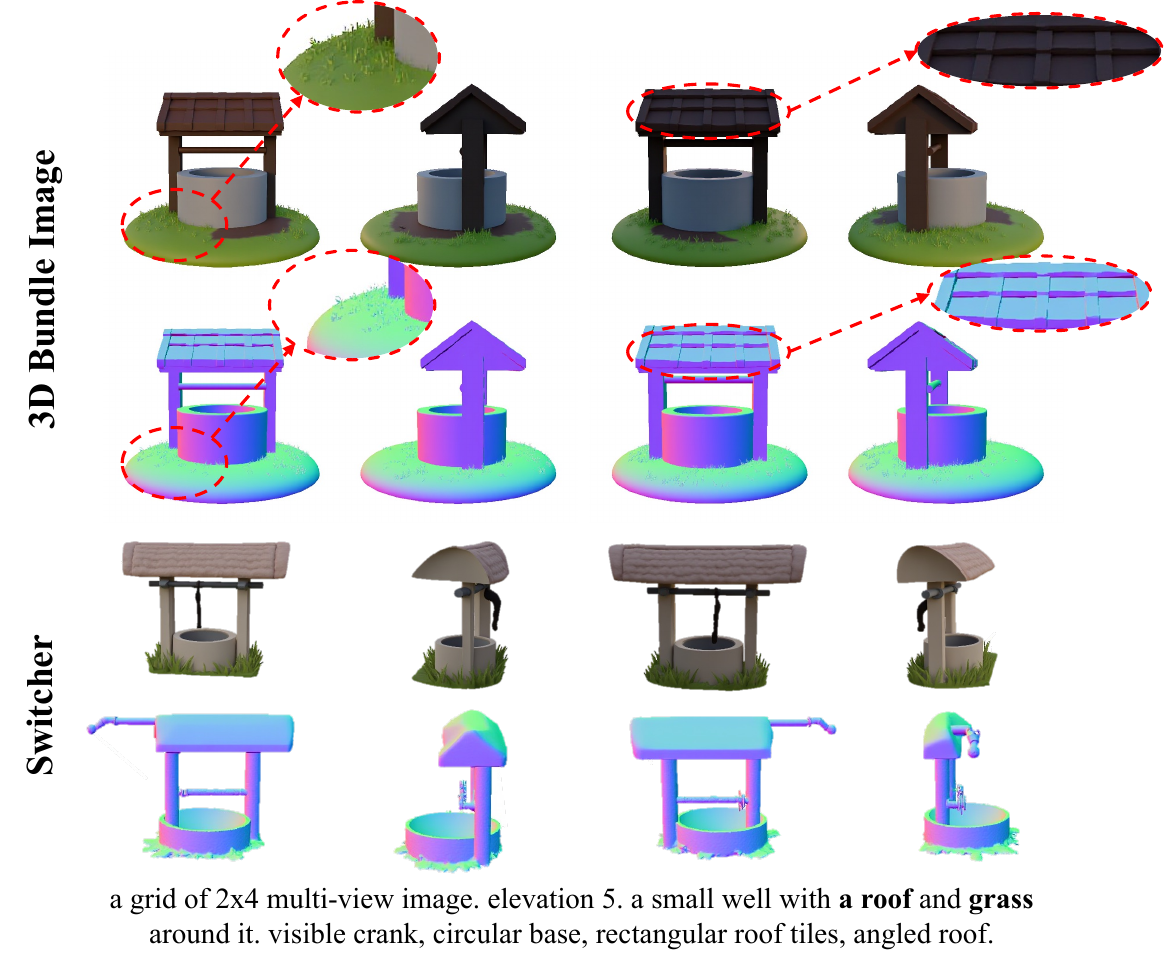}
    \vspace{-1.8em}
    \caption{Ablating the mechanisms of generating multiview RGB and normal maps. Both our ``3D Bundle Image'' and  ``Switcher''~\cite{long2023wonder3d,li2024era3d} are built upon Flux.1-dev~\cite{blackforest2024flux} model. }
    \label{fig:ablation_switcher}
    \vspace{-1em}

\end{figure}

\vspace{-0.1cm}
\noindent\textbf{The mechanism of 3D Bundle Image} As discussed earlier, we propose a novel approach of combining RGB and normal maps into a single image, referred to as the ``3D Bundle Image," for model training. We compare this approach with the ``Switcher" mechanism used in prior works \cite{long2023wonder3d,li2024era3d}. Unlike our model, which generates RGB images and normal maps concurrently, these works employ a ``Switcher" to selectively produce either RGB images or normal maps. As illustrated in Fig.~\ref{fig:ablation_switcher}, the ``Switcher" mechanism fails to maintain coherence between the two outputs. In contrast, our ``3D Bundle Image" achieves significantly higher consistency. This improvement is due to the DiT model's architecture, particularly its attention blocks, which effectively capture long-range dependencies and interactions. However the ``Switcher" mechanism processes the modalities separately and does not leverage this advantage. 

% This improvement can be attributed to the its effective utilization of the DiT model’s architecture, particularly its attention blocks, which are designed to capture long-range dependencies and interactions. By capitalizing on the structural advantages of DiT, the 3D Bundle Image input promotes consistent and accurate representations across both RGB and normal maps, resulting in coherent multi-view and cross-modal relationships.

% \vspace{-0.2cm}
\noindent\textbf{Dataset scale}
% 我们有5k、10k、50k、100k、140k五个不同的模型
% 标记不同工作的数据集
% unique3d: 50k, LGM: 80K
% Direct2.5 500K 
% 3DTopia: 320K 3D objects for 118k iter, fine-tune 70k high quality 26k
In Sec.~\ref{sec:experiments}, we compare the Kiss3DGen-Base model with baseline models, demonstrating that it achieves state-of-the-art performance. However, we note that the scale of datasets varies significantly across different works. 
For instance, 3DTopia~\cite{hong20243dtopia} was trained on over 320k 3D objects, Direct2.5~\cite{lu2024direct2} used 500k, and Unique3D~\cite{wu2024unique3d} employed only 50k objects.
We reduce our dataset to 50k objects and train a model named Kiss3DGen-50k, and find that the model still performs well in most tasks (Tab.~\ref{tab:text-to-3D}), proving the effectiveness of our framework.
In comparison, we find that the Kiss3DGen model trained on a small scale of data sometimes failed to generate a 3D Bundle Image. See more details in our supplement.
\vspace{-0.15cm}
\section{Concluding Remarks}
\label{sec:conclusion}
We introduces Kiss3DGen, a straightforward yet highly effective approach for a variety of 3D generation tasks. By leveraging knowledge from pretrained 2D diffusion models, it seamlessly integrates with existing techniques like ControlNet. Despite its simplicity, Kiss3DGen excels across tasks such as text-to-3D, image-to-3D, 3D enhancement, and 3D editing. It also demonstrates strong performance even with limited training data, benefiting from the preservation of knowledge in Flux. Additionally, the model generalizes well, enabling the generation of objects beyond its original 3D training set.

There is significant potential for further improvement, particularly in exploring the optimal representation of geometry (e.g., normal maps) and developing efficient methods for generating high-resolution views. These challenges will be addressed in our future work.
{
    \small
    \bibliographystyle{ieeenat_fullname}
    \bibliography{main}
}


\end{document}

\clearpage
\setcounter{page}{1}
\maketitlesupplementary

The supplementary file includes a demo video showcasing the performance of our model in various tasks, including text/image-to-3D generation, and 3D enhancement/editing. Additionally, we provide further studies and detailed explanations below to offer a deeper understanding of the model and its capabilities. We will release the code upon acceptance. 

\section{Ablating the Initialization of Mesh}
\begin{figure}[h]
    \centering
    \includegraphics[width=1.0\linewidth]{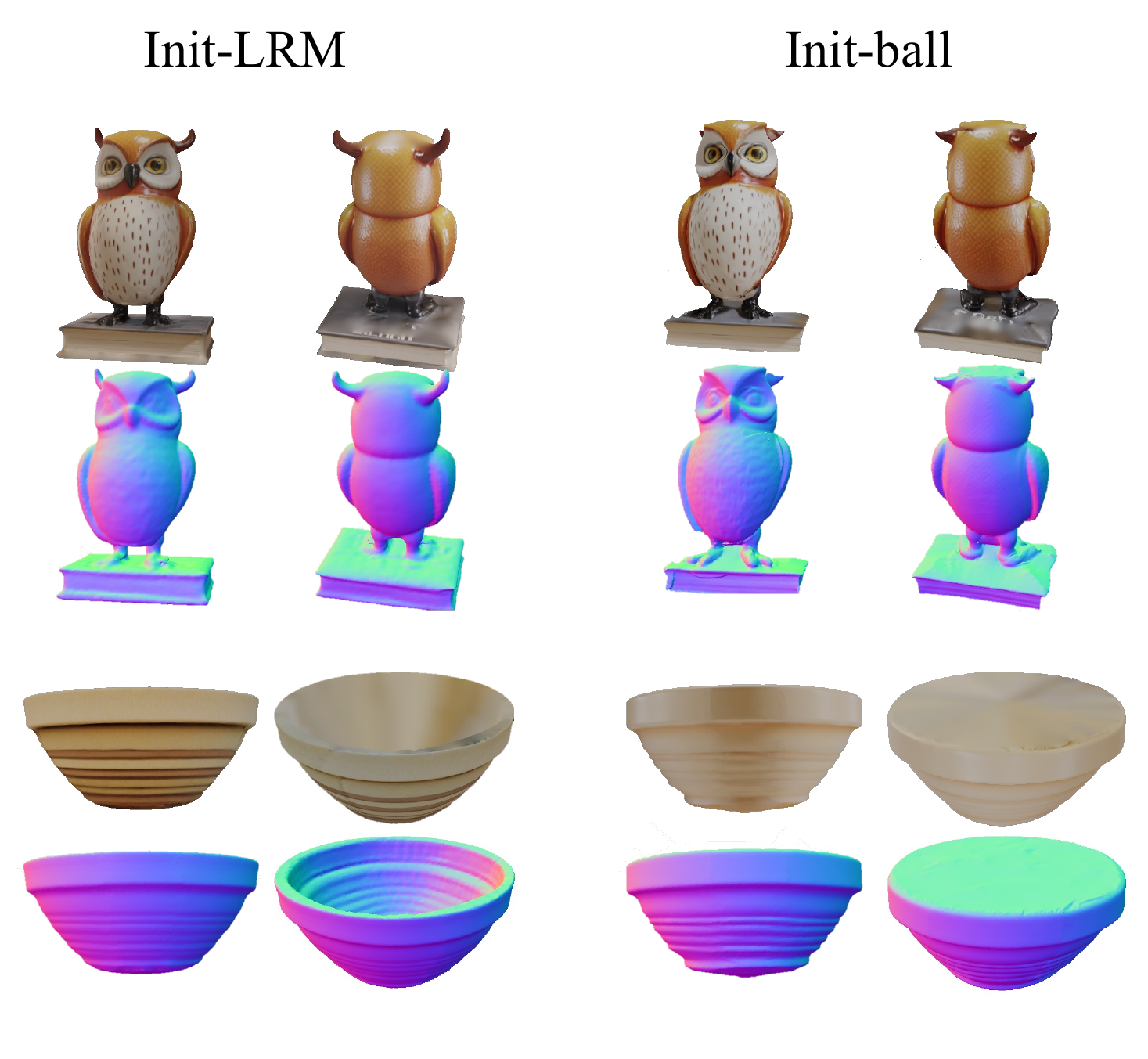}
    \vspace{-2em}
    \caption{Qualitative comparison of 3D reconstruction results between different initializations in the reconstruction stage of our framework. The upper case (owl) shows that using LRM or sphere initialization yields similar results. The second row (bowl) shows that using sphere initialization may fail at capturing the concave geometric structure, while using LRM mitigates this problem. }
    \label{fig:ablation_init}
\end{figure}

In our manuscript, we adopt the off-the-shelf LRM~\cite{ge2024prm} model or a simple sphere shape to initialize the coarse mesh, then refine the mesh with ISOMER~\cite{wu2024unique3d}. We have also experimented with different settings, such as refining the mesh from a simple, sphere-shaped initialization. As shown in Fig.~\ref{fig:ablation_init}, the results are still of excellent overall quality; however, there appear to be more geometrical errors at unseen surfaces. 
We also conducted quantitative evaluations, as shown in Table~\ref{tab:text_to_3d_eval} and Table~\ref{tab:image_to_3d_eval}. The quantitative results demonstrate that the LRM initialization generally outperforms the sphere initialization across most metrics.

% Text-to-3D Table
\begin{table}[!ht]
    \caption{Quantitative comparison of generated results for \textbf{text-to-3D} with different initializations at the reconstruction stage.}
    \label{tab:text_to_3d_eval}
    \centering
    \resizebox{0.9\columnwidth}{!}{
    \begin{tabular}{cccc}
    \hline
    Method     & CLIP$\uparrow$ & Quality$\uparrow$ & Aesthetic$\uparrow$ \\ \hline
    Init-LRM   & 0.837          & 2.700             & 1.800               \\ 
    Init-Sphere & 0.8012         & 2.559             & 1.566               \\ \hline
    \end{tabular}
    }
\end{table}

% Image-to-3D Table
\begin{table}[!ht]
    \caption{Quantitative comparison of generated results for \textbf{image-to-3D} with different initializations at the reconstruction stage.}
    \label{tab:image_to_3d_eval}
    \centering
    \resizebox{1.0\columnwidth}{!}{
        \begin{tabular}{cccccc}
            \hline
            Method      & CD$\downarrow$ & FS$\uparrow$ & PSNR$\uparrow$ & SSIM$\uparrow$ & LPIPS$\downarrow$ \\ \hline
            Init-LRM    & 0.149          & 0.769       & 20.348         & 0.902          & 0.116             \\ 
            Init-Sphere & 0.173          & 0.719       & 20.122         & 0.902          & 0.117             \\ \hline
        \end{tabular}
    }
\end{table}

\begin{figure}[ht]
    \centering
    \includegraphics[width=0.5\textwidth]{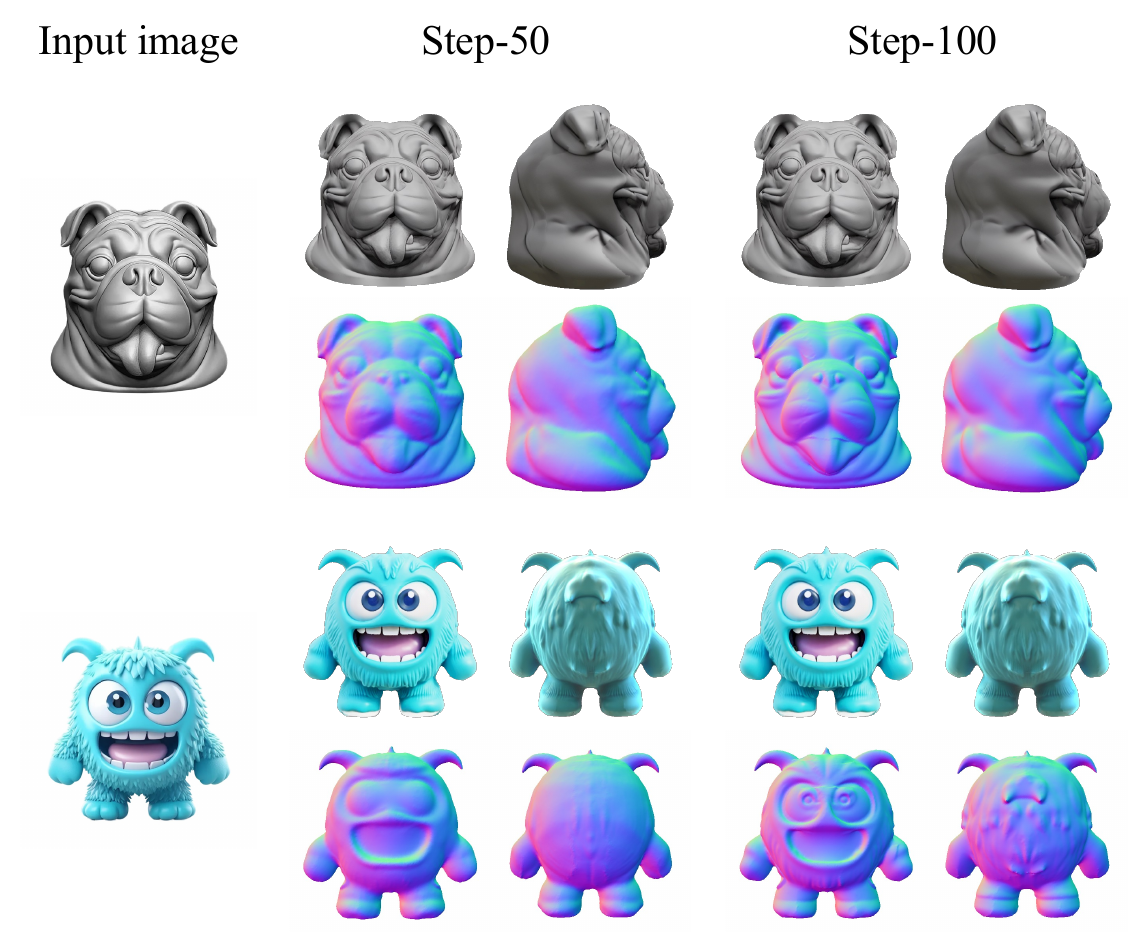}
    \caption{Qualitative comparison of 3D reconstruction results with different optimization steps with ISOMER~\cite{wu2024unique3d}. As shown, optimizing with more steps leads to finer geometrical details. }
    \label{fig:ablation_step}
\end{figure}

\section{Ablating the number of steps in ISOMER}
In our main manuscript, we proposed using the off-the-shelf LRM~\cite{xu2024instantmesh, hong2024lrmlargereconstructionmodel} model to initialize the coarse mesh, followed by ISOMER~\cite{wu2024unique3d} to optimize and produce the final mesh. In the optimization step, there is a critical parameter that controls the number of geometry optimization steps. This parameter directly impacts the inference time. Specifically, when the number of steps is set to 50, the geometry optimization step takes approximately 5 seconds, while setting it to 100 increases the time to about 10 seconds.
To understand the effect of this parameter, we conducted an ablation study, as shown in Fig.~\ref{fig:ablation_step}. The results indicate that increasing the number of steps leads to sharper and more refined geometry, albeit at the cost of longer computation time. It is worth noting that, in our main manuscript, we used a step value of 50 for all experiments to balance experimental efficiency and result quality.
This analysis highlights the trade-off between optimization time and geometry refinement, providing guidance for parameter selection based on application requirements.

\section{Compatibility and extensibility of methods.} 
\begin{figure}[]
    \centering
    \includegraphics[width=0.9\linewidth]{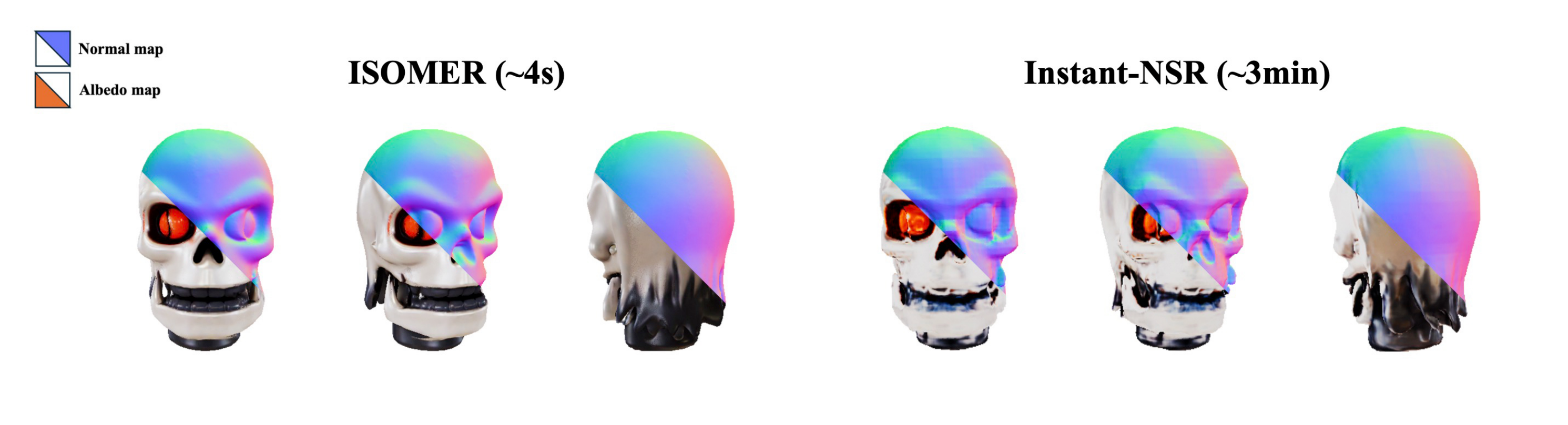}
    % \vspace{-1.8em}
    \caption{\textit{{Visual comparisons of different reconstruction methods.}}}
    % \vspace{-1.0em}
    \label{fig:recon_comparison}
\end{figure}

\begin{figure}[]
    \centering
    \includegraphics[width=0.7\linewidth]{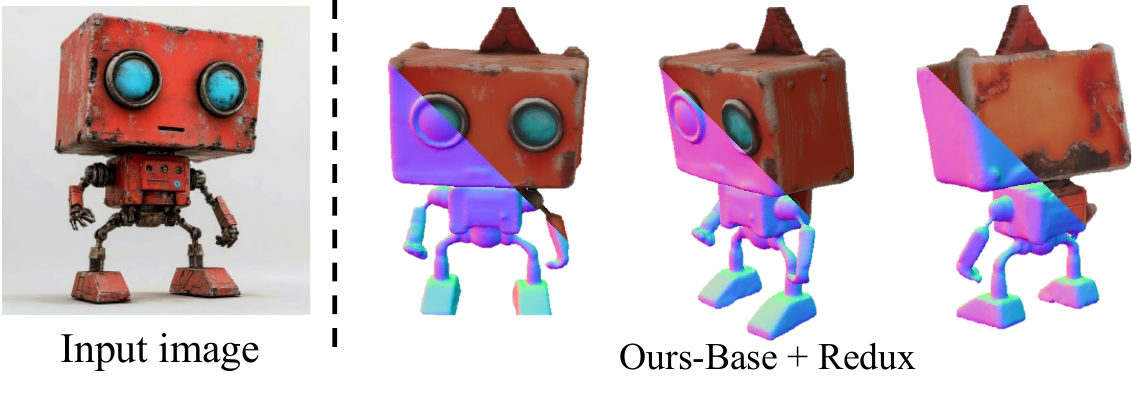}
    % \vspace{-1.2em}
    \caption{\textit{{Visualization of \textcolor{blue}{image-to-3d} with \textcolor{blue}{redux}}}.}
    % \vspace{-2.0em}
    \label{fig:redux}
\end{figure}

As shown in Fig.~\ref{fig:recon_comparison}, our method is compatible with reconstruction techniques besides ISOMER, such as Instant-NSR. Additionally, our approach retains DiT’s full capabilities, enabling seamless integration with tools like IP-Adapter, ControlNet, or Flux Redux \footnote{\url{https://blackforestlabs.ai/flux-1-tools}} (Fig.~\ref{fig:redux}), highlighting its adaptability and extensibility.

\section{System efficiency.} In Tab.~\ref{tab:inference_time}, we quantitatively measure the inference time of our framework and baseline methods on an A800 GPU, our approach achieves the best performance within reasonable inference time.

\section{More qualitative comparisons.} 
\begin{figure}[]
    \centering
    \includegraphics[width=\linewidth]{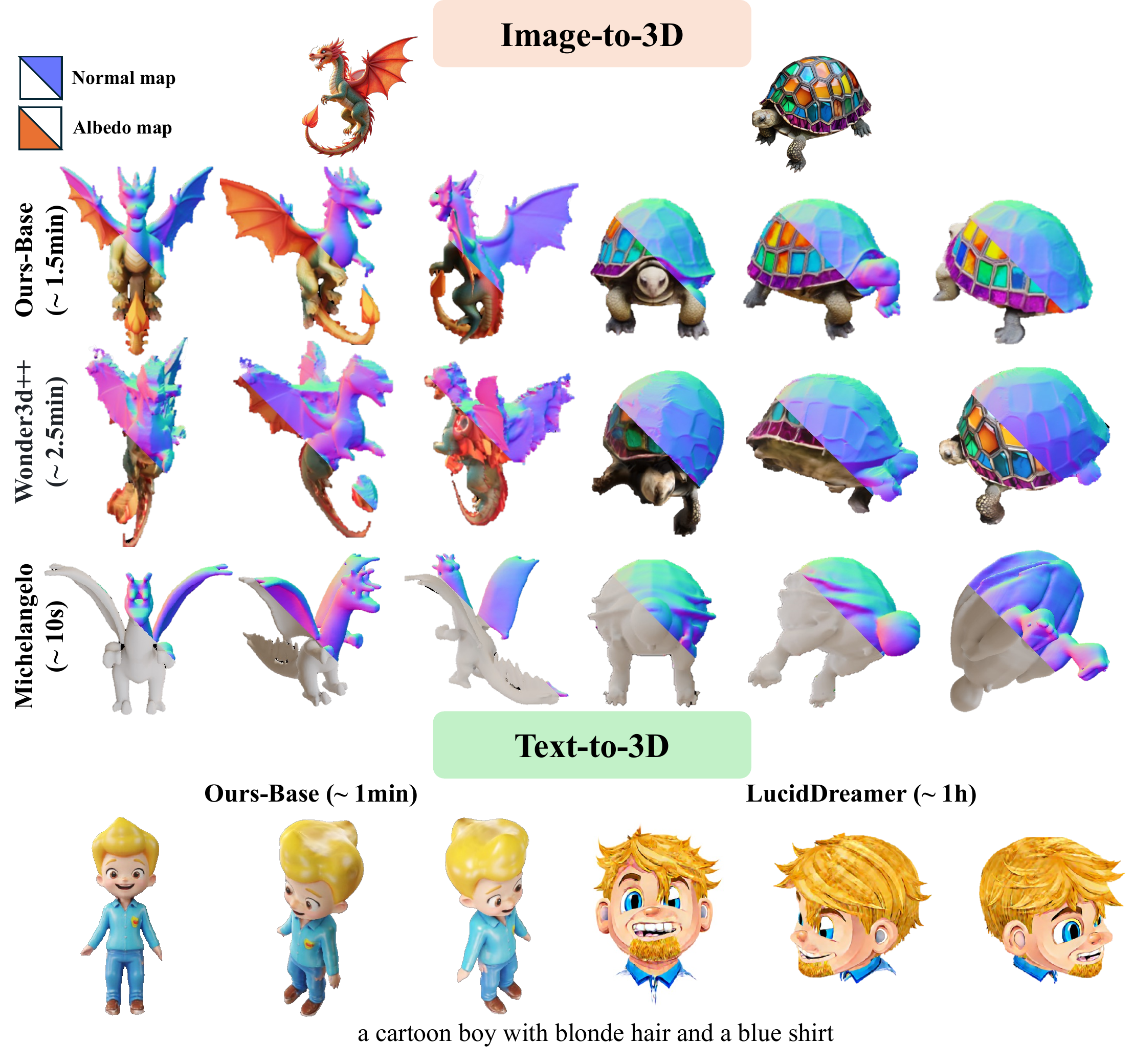}
    \vspace{-2.2em}
    \caption{\textit{{Qualitative comparisons with more state-of-the-art methods for \textcolor{blue}{image-to-3D} and \textcolor{blue}{text-to-3D} generation.}}}
    \vspace{-0.5em}
    \label{fig:image_to_3D}
\end{figure}

We demonstrate more comparisons against Wonder3D++ and Michelangelo for image-to-3D and LucidDreamer for text-to-3D in Fig.~\ref{fig:image_to_3D}. Our method achieves better results in texture details, semantic alignment, and text-3D consistency.

\begin{table}[]
    \caption{Comparison of inference time with other methods in different tasks. (in seconds). ``--'' means unapplicable.}
    \label{tab:inference_time}
    \centering
    \vspace{-1em}
    \resizebox{\linewidth}{!}{
    \begin{tabular}{cccccccc}
    \hline
        Task          & ours    & MVEdit   & Hunyuan3D-1.0  & CraftsMan & Unique3D & 3DTopia & Direct2.5  \\ \hline
        Text-to-3D    & 56.8   & --       & 105.0            & --        & --       & 240.0     & 163.6     \\
        Image-to-3D   & 87.3   & --       & 79.9          & 6.0         & 37.2    & --      & --         \\
        3D-to-3D      & 71.7   & 360.0   & --             & --        & --       & --      & --         \\ \hline
    \end{tabular}
    }
    \vspace{-0.5em}
\end{table}

\section{User Study}
In our manuscript, we conduct quantitative evaluations comparing our method with baseline methods, demonstrating its superior performance. We also present a user study to assess user preferences.

The user study was conducted on Amazon Mechanical Turk\footnote{\url{https://www.mturk.com}}, involving 180 participants. To ensure quality, we included attention-check questions to filter out inattentive responses, resulting in 80 qualified participants whose responses were analyzed. Ultimately, we collected 2,000 valid responses covering key aspects such as geometry quality, texture quality, and overall quality. The results used in user study are generated with the default hyper-parameters without any cherry-picking.

\begin{figure}[t]
    \centering
    \includegraphics[width=1.0\linewidth]{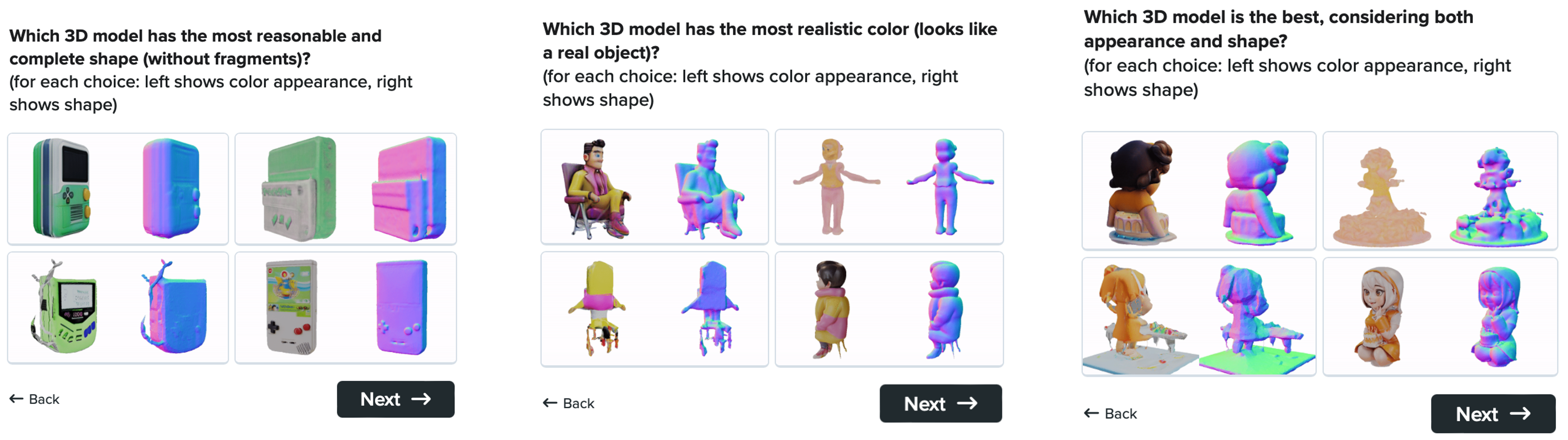}
    \caption{Screenshots of our user study questionnaire.}
    \label{fig:questionnaire}
\end{figure}

Figure~\ref{fig:questionnaire} shows a screenshot of the user study questionnaire. The options included GIFs displaying orbital views of the object in both color and normal space, allowing participants to better visualize the 3D structure and texture, thereby enhancing their ability to provide informed feedback.

For each case in the user study, we present a video to the users where the object rotates 360 degrees, with the left side displaying the RGB map and the right side showing the Normal map. Users are asked to select the best result based on a series of questions. In terms of question design, we focus on several key aspects:
\begin{enumerate}
    \item \textbf{Geometry}: "Which 3D model has the most reasonable and complete shape (without fragments)?"
    % , "Which 3D model has the most reasonable and detailed shape?"
    \item \textbf{Texture}: "Which 3D model has the most realistic color (looks like a real object)?"
    % \item \textbf{Alignment}: "Which 3D model has appearance that best matches the above image?",  "Which 3D model has appearance that best matches the text description below?" 
    \item \textbf{Overall quality}: "Which 3D model is the best, considering both appearance and shape?"
\end{enumerate}

The results of the user study are summarized in Tab.~\ref{tab:userstudy}, where it can be observed that our method outperforms the baselines in terms of user preference for both geometry and texture quality, as well as overall impression.

\begin{table}[h!]
\centering
\caption{Study on user's preference on 3D generation results of ours and baseline methods.}
\label{tab:userstudy}
\begin{tabular}{ccc}
\hline
\textbf{Category} & \textbf{Method} & \textbf{Percentage} \\ \hline
\textbf{Texture}  & Ours            & 35.47\%             \\
                  & Hunyuan         & 32.37\%             \\ 
                  & Unique3D        & 13.13\%             \\ 
                  & 3Dtopia         & 6.75\%              \\ 
                  & Direct2.5D      & 12.28\%             \\ \hline
\textbf{Geometry}    & Ours            & 37.61\%             \\ 
                  & Hunyuan         & 36.24\%             \\ 
                  & Unique3D        & 10.45\%             \\ 
                  & 3Dtopia         & 5.13\%              \\ 
                  & Direct2.5D      & 10.56\%             \\ \hline
\textbf{Overall Quality}  & Ours            & 38.72\%             \\ 
                  & Hunyuan         & 32.18\%             \\
                  & Unique3D        & 15.04\%             \\ 
                  & 3Dtopia         & 6.49\%              \\ 
                  & Direct2.5D      & 7.57\%              \\ \hline
\end{tabular}
\end{table}

\section{Applications and visualization}
\begin{figure*}[h]
    \centering
    \includegraphics[width=0.72\linewidth]{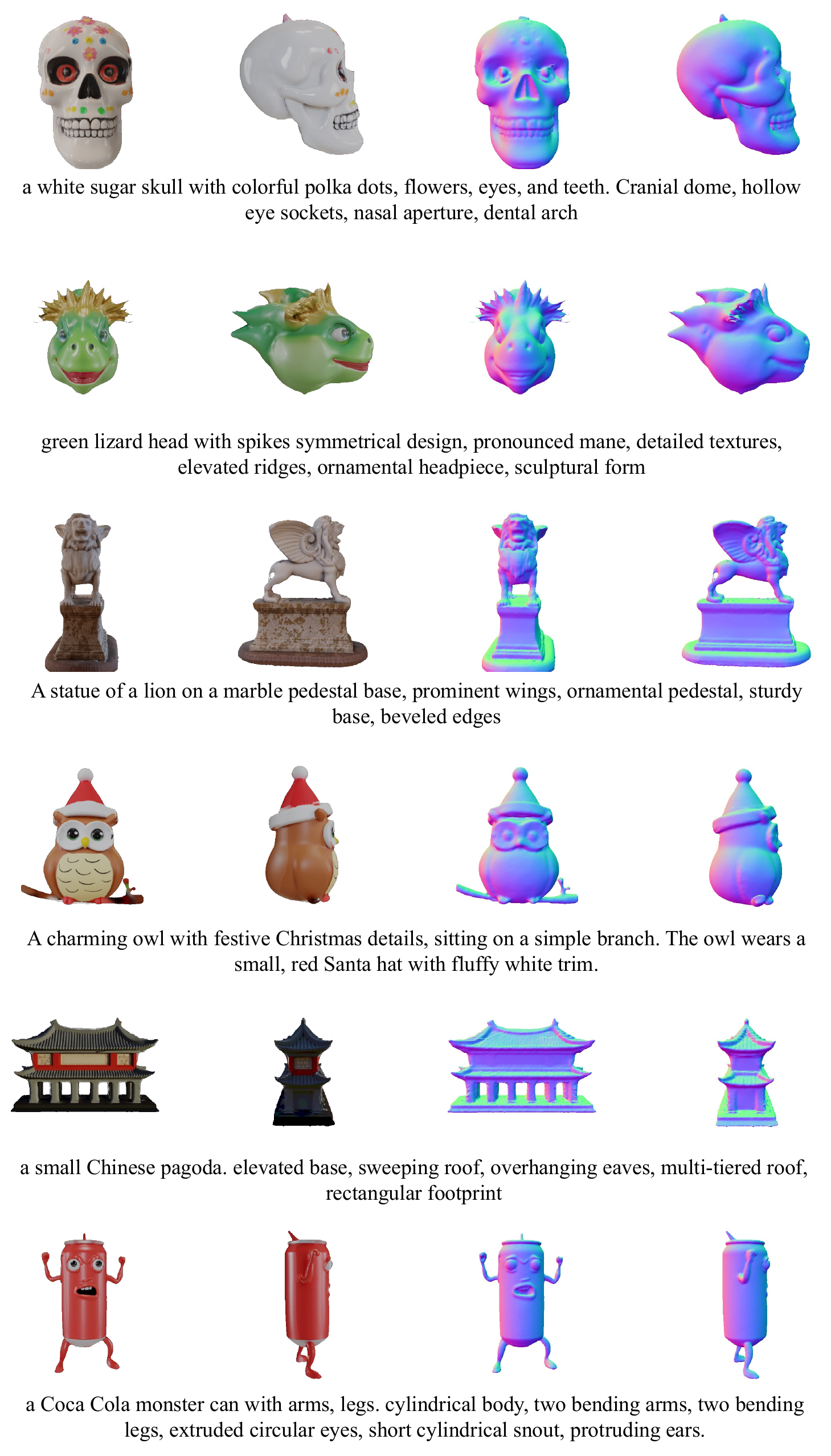}
    \caption{More show cases of \textbf{Text-to-3D} generation with our model. Please zoom in for details.}
    \label{fig:T23D_more_result}

\end{figure*}

\begin{figure*}[h]
    \centering
    \includegraphics[width=0.72\linewidth]{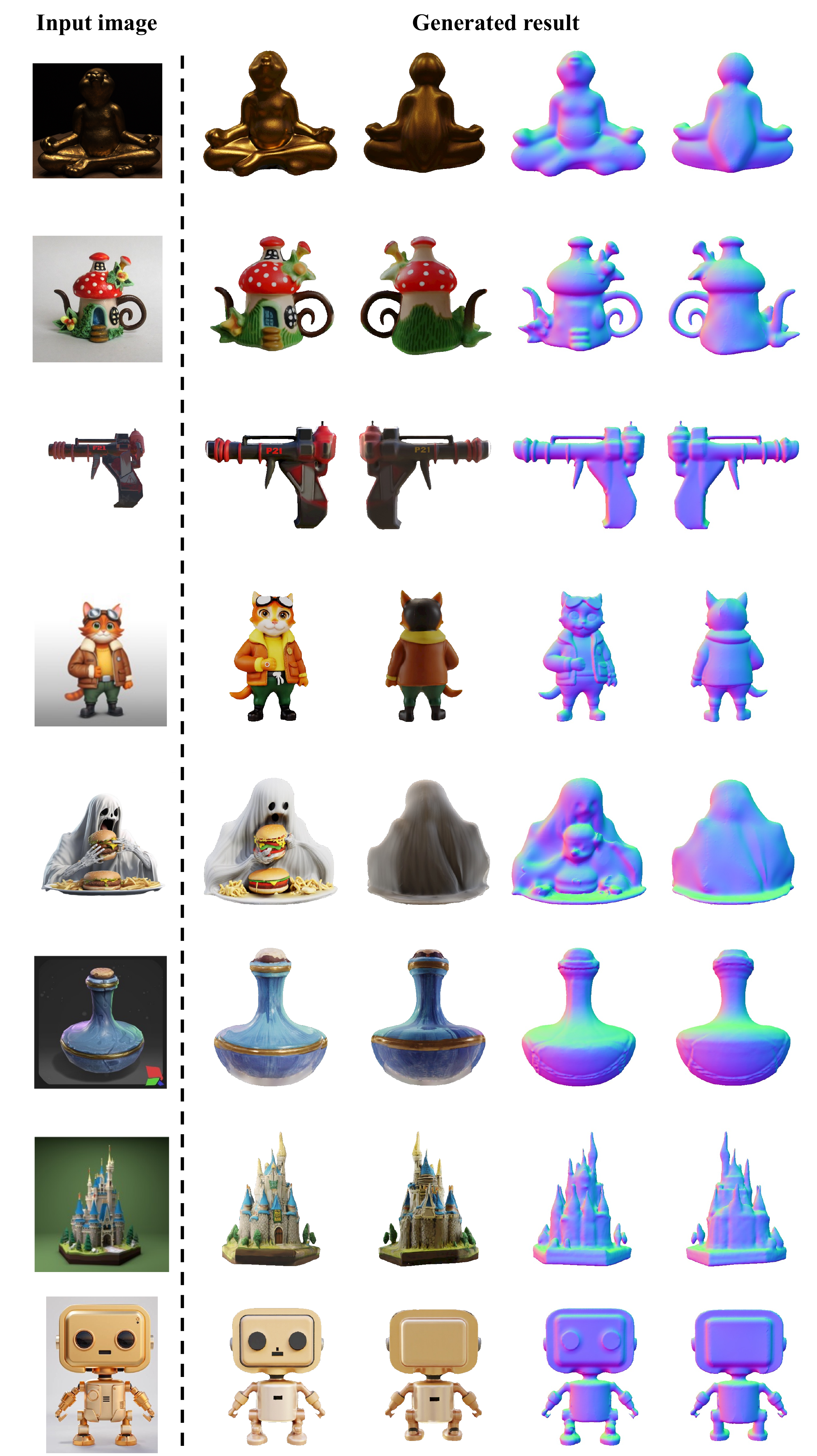}
    \caption{More show cases of \textbf{Image-to-3D} generation with our model. Please zoom in for details.}
    \label{fig:Img23D_more_result}

\end{figure*}

\begin{figure*}[h]
    \centering
    \includegraphics[trim=30 0 30 0, width=1.0\linewidth]{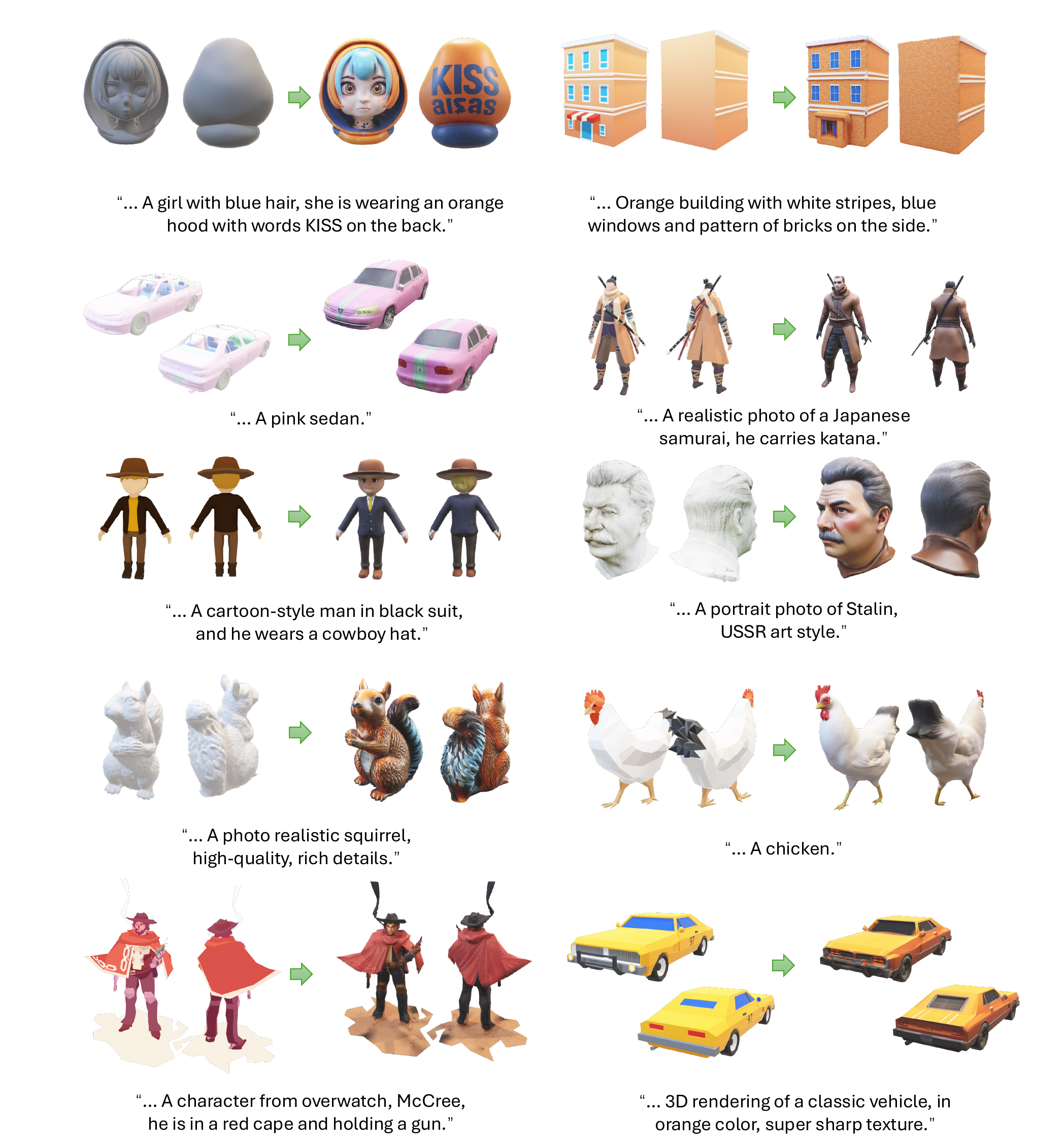}
    \caption{\textbf{3D enhancement and editing results with our model.} Notably, we adopt off-the-shelf controlNets~\cite{zhang2023adding}, e.g, Normal, Canny and Tile, with our Kiss3DGen model to align the generation results with the input 3D models. For simplicity, we denote the fixed camera control caption as ``...", and the detailed-object captions are manually crafted to achieve desired results. Please zoom in for details.}
    \label{fig:supp-3d-edit}
\end{figure*}

In our main paper, we introduce various applications with our model, including text-to-3d, image-to-3d, 3D editing and enhancement.
We demonstrate more results in Fig.~\ref{fig:T23D_more_result}, Fig~\ref{fig:Img23D_more_result} and Fig.~\ref{fig:supp-3d-edit}. 
Also, we attach a video to this supplementary to present the 3D generation results in a dynamic approach.

\begin{figure}[t]
    \centering
    \includegraphics[trim=40 0 10 0, width=1.0\linewidth]{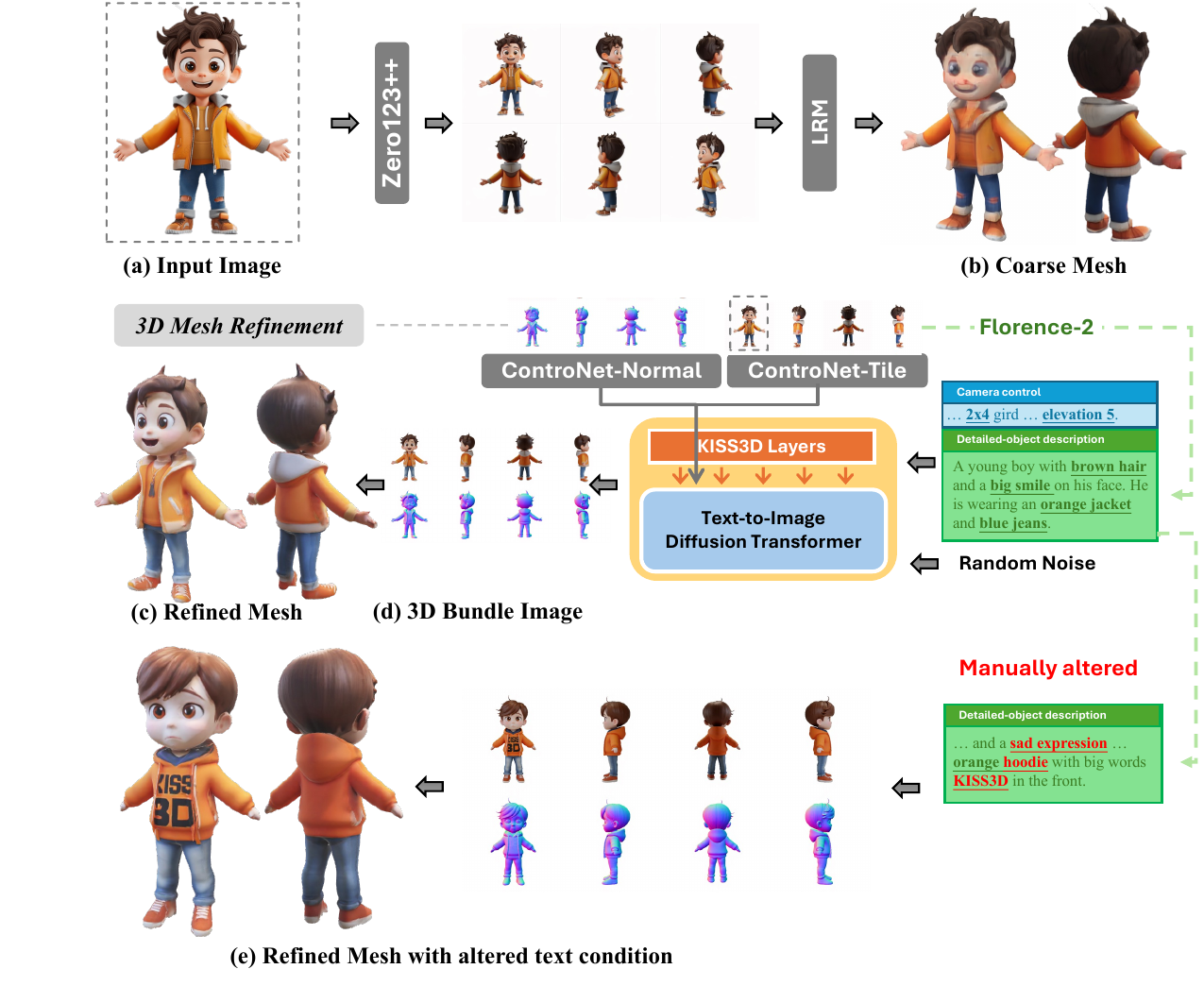}
    \caption{\textbf{Advanced image-to-3D pipeline with our framework.} In this case, we alter the text descriptions at the 3D mesh refinement stage and achieve accurate textual control on the refined result. Please zoom in for details.}
    \label{fig:advanced-img-to-3d}
\end{figure}

\subsection{Advanced image to 3D}
In Fig.~\ref{fig:advanced-img-to-3d}, we illustrate a 3D generation pipeline that utilizes multi-modal conditions, including both images and text. Unlike most existing image-to-3D generation methods that produce 3D assets aligned solely with the input image, our framework introduces textual control over the generation outcomes, significantly enhancing the utility of 3D content creation from images. This capability allows for more nuanced and tailored 3D outputs, catering to specific user requirements.
And notably, the application of our model extends beyond the examples presented in this paper.

\section{Limitations}
In this paper, we effectively adapt the pretrained 2D diffusion transformer model, specifically Flux~\cite{blackforest2024flux}, for the generation of 3D Bundle Images. To maximize the potential of the Flux model, we render our 3D dataset under varying environmental illuminations, enhancing its similarity to real-world images on which the Flux model was trained.
As a result, the generated 3D Bundle Image retains lighting information, which was not disentangled from the model texture during the reconstruction phase of this work. We leave this for future study.

{
    \small
    \bibliographystyle{ieeenat_fullname}
    \bibliography{main}
}
